\documentclass[conference]{IEEEtran}
\usepackage{amsmath}

\usepackage{amssymb,amsfonts}
\usepackage{algorithm}
\usepackage{algpseudocode}
\usepackage{xspace}
\usepackage{booktabs}
\usepackage{pifont} %
\usepackage{graphicx}
\usepackage{textcomp}
\usepackage{xcolor}
\usepackage{subcaption}
\usepackage{fancyhdr}
\usepackage{hyperref}
\usepackage{xspace}
\usepackage{enumitem}
\usepackage{booktabs}
\usepackage{colortbl}
\usepackage{multirow}
\usepackage{makecell}
\usepackage[capitalise]{cleveref}
\usepackage{soul,color}
\usepackage{multicol}

\algnewcommand\algorithmicinput{\textbf{Input:}}
\algnewcommand\algorithmicoutput{\textbf{Output:}}
\algnewcommand\Input{\item[\algorithmicinput]}
\algnewcommand\Output{\item[\algorithmicoutput]}

\newcommand{\Rebuttal}[1]{{\color{black}#1}}

\newcommand{\methodnamebold}{\textbf{\textit{Gossamer}}\xspace}
\newcommand{\methodname}{{Gossamer}\xspace}
\newcommand{\methodnameslim}{{Gossamer}\xspace}
\newcommand{\x}{$\times$\xspace}

\definecolor{circlecolor}{HTML}{6D9CF4}
\usepackage{tikz}
\newcommand{\bluecircle}[1]{%
  \tikz[baseline=(char.base)]{
    \node[
      circle,
      fill=circlecolor,
      inner sep=0.6pt,
      minimum size=10pt
    ](char){\scriptsize\textbf{\textcolor{white}{#1}}};
  }%
}

\newcounter{rebuttallabelcounter}

\IEEEoverridecommandlockouts
\def\BibTeX{{\rm B\kern-.05em{\sc i\kern-.025em b}\kern-.08em
    T\kern-.1667em\lower.7ex\hbox{E}\kern-.125emX}}
\begin{document}

\title{
Gossamer: A Utility-Driven Architecture for Constant-Budget KV Cache Compression in Reasoning LLMs
}

\author{
\IEEEauthorblockN{
Feng Cheng\IEEEauthorrefmark{2},
Cong Guo\IEEEauthorrefmark{2}\IEEEauthorrefmark{4},
Junyao Zhang\IEEEauthorrefmark{2},
Haoxuan Shan\IEEEauthorrefmark{2},
Chiyue Wei\IEEEauthorrefmark{2},
Hong Wang\IEEEauthorrefmark{3},
Hai Li\IEEEauthorrefmark{2},
Yiran Chen\IEEEauthorrefmark{2}}
\IEEEauthorblockA{\IEEEauthorrefmark{2}\textit{Duke University}, Durham, NC, USA\\
\{feng.cheng, cong.guo, jz420, haoxuan.shan, chiyue.wei, hai.li, yiran.chen\}@duke.edu}
\IEEEauthorblockA{\IEEEauthorrefmark{3}\textit{Intel Corporation}, Santa Clara, CA, USA\\
hong.wang@intel.com}
\IEEEauthorblockA{\IEEEauthorrefmark{4}Corresponding author}
}

\maketitle

\begin{abstract}
Transformer-based large language models (LLMs) are increasingly applied to complex reasoning tasks, yet their inference efficiency is fundamentally constrained by the rapid growth of the KV cache.  
As reasoning contexts extend to hundreds of thousands of tokens, the key-value (KV) cache becomes the dominant contributor to both memory footprint and bandwidth demand, tightly coupling computation with memory utilization and thereby limiting scalability.  
Existing compression methods, such as quantization and eviction, alleviate capacity pressure but remain purely algorithmic, often introducing additional passes and compute overhead that reduce system efficiency.  

This paper presents \methodname, a KV cache compression architecture for reasoning LLMs that bridges the algorithm--system--hardware gap.  
\methodname introduces a unified utility-based action framework that dynamically selects compression operations, quantization, merging, or eviction, under a constant KV cache budget.  
Through cross-domain co-design, we fuse cosine-similarity estimation directly into the attention computation, exploiting idle compute cycles to perform similarity evaluation without latency or bandwidth cost. We call this mechanism \textit{Free-Lunch Utilization}.  
This enables real-time, fine-grained KV management with near-constant memory usage and negligible overhead.  
Under tight memory constraints, \methodname sustains reasoning quality comparable to uncompressed inference while improving end-to-end throughput by up to 16.7\x over uncompressed baselines, 5.5\x and 4.9\x over the quantization-based Oaken and Ecco, and 1.8\x over the eviction-based R-KV.   
\end{abstract}


\section{Introduction}
\label{sec:introduction}

Transformer-based large language models (LLMs) \cite{OpenAI_ChatGPT_2022,touvron2023llama1,jiang2023mistral_7b,bai2023qwen} have emerged as powerful reasoning engines, enabling complex decision-making, planning, and multi-step generation across diverse domains.  
As model size and reasoning context length increase, the computational and memory requirements of inference grow rapidly~\cite{wang2024sofa,lin2024infinite,pope2023efficiently,cerab}, posing serious challenges for deploying reasoning LLMs efficiently under limited hardware resources.  
Modern reasoning LLMs employ a \textbf{key-value (KV) cache}~\cite{vaswani2017attention,pope2023efficiently,wang2025kvcache} to accelerate autoregressive decoding.  
The KV cache stores attention key and value tensors from previous steps, allowing the model to reuse them for subsequent token generation instead of recomputing all past activations.  
While this mechanism is essential for long-form reasoning, its linearly growing size creates a substantial memory footprint and introduces severe memory bandwidth bottlenecks.

From the memory perspective, the KV cache expansion directly increases storage demand and bandwidth pressure.  
The size of the KV cache grows linearly with the reasoning context and model dimension, quickly becoming the dominant contributor to runtime memory consumption~\cite{wang2024sofa}.  
For short contexts (e.g., 512 tokens), the KV cache accounts for less than 2\% of total memory usage, but at long contexts (e.g., 128K tokens), it can occupy over 84\% of available capacity~\cite{hooper2024kvquant}.  
This growth stresses on-device memory and amplifies memory traffic, as cached data must be repeatedly accessed during each decoding step.  
When the KV cache exceeds available capacity, systems often resort to \emph{offloading} strategies~\cite{sheng2023flexgen,zhang2023h2o,long2025sliminfer} that migrate cache data across the memory hierarchy or to external storage, introducing additional data movement and latency.  

From the computation perspective, 
a larger KV cache directly increases the arithmetic cost of attention, since each decoding step must process more KV entries. 
At the same time, limited memory reduces batching opportunities~\cite{kwon2023efficient_vllm}, leading to low compute utilization and degraded throughput.  
This tight {computation and memory coupling} fundamentally limits inference efficiency for reasoning workloads.  
Consequently, there is a critical need for architectural techniques that reduce the KV cache footprint while maintaining reasoning quality.

A natural response to the growth of the KV cache is compression.  
Existing works~\cite{rkv,lin2024qserve,pyramidkv,zhang2405kv,hooper2024kvquant,li2025commvq,zhang2023h2o,cheng2025ecco,li2024snapkv,yuan2024kv,kelle2025,chang2025palu,archsurvey,prosperity,wei2026focus,transarr,phi,mx2fp} explore several directions, including quantization, pruning, and eviction-based methods. These broadly fall into two categories.  
The first category is \textbf{local in-place compression}~\cite{cheng2025ecco,lin2024qserve,fu2024lazyllm,hooper2024kvquant,platinum}.  
These methods compress the KV cache by a fixed ratio, such as quantizing from FP16 to 4-bit~\cite{lin2024qserve} or pruning a constant fraction of tokens~\cite{fu2024lazyllm}.
Such techniques delay memory exhaustion but do not fundamentally solve the problem, as the KV cache still grows linearly with context length.  
For efficient inference, they rely only on local information and make per-token decisions without global context awareness.  

The second category leverages \textbf{global information} from previous reasoning history to guide compression.  
In reasoning LLMs, many cached tokens are redundant across multi-turn reasoning steps~\cite{wang2025kvcache}.  
Recent global-aware methods~\cite{li2024snapkv,pyramidkv,rkv,wang2025kvcache,wang2024model,zhang2024cam} compare new KV entries with the existing KV cache to decide whether to retain or evict them, effectively bounding the KV cache within a constant budget, such as 512 entries.  
However, these methods emphasize algorithmic effectiveness and often overlook architectural efficiency.  

Global-aware compression requires computing similarities such as cosine distance across the entire cache~\cite{rkv}, which is computationally expensive.  
Our measurements show that a single global compression step on a 512-entry cache can take up to twelve times the latency of one attention computation.  
To amortize this cost, these methods commonly introduce a secondary buffer and accumulate new tokens, for example, 128 at a time, before performing global updates.  
While this improves amortized efficiency, it sacrifices responsiveness and does not align with real-time inference pipelines, leading to suboptimal accuracy.
As a result, current KV cache compression methods remain algorithmically driven but architecturally inefficient, leaving a clear gap between compression utility and hardware efficiency.

\begin{figure}[t]
  \centering
  \includegraphics[width=0.95\linewidth]{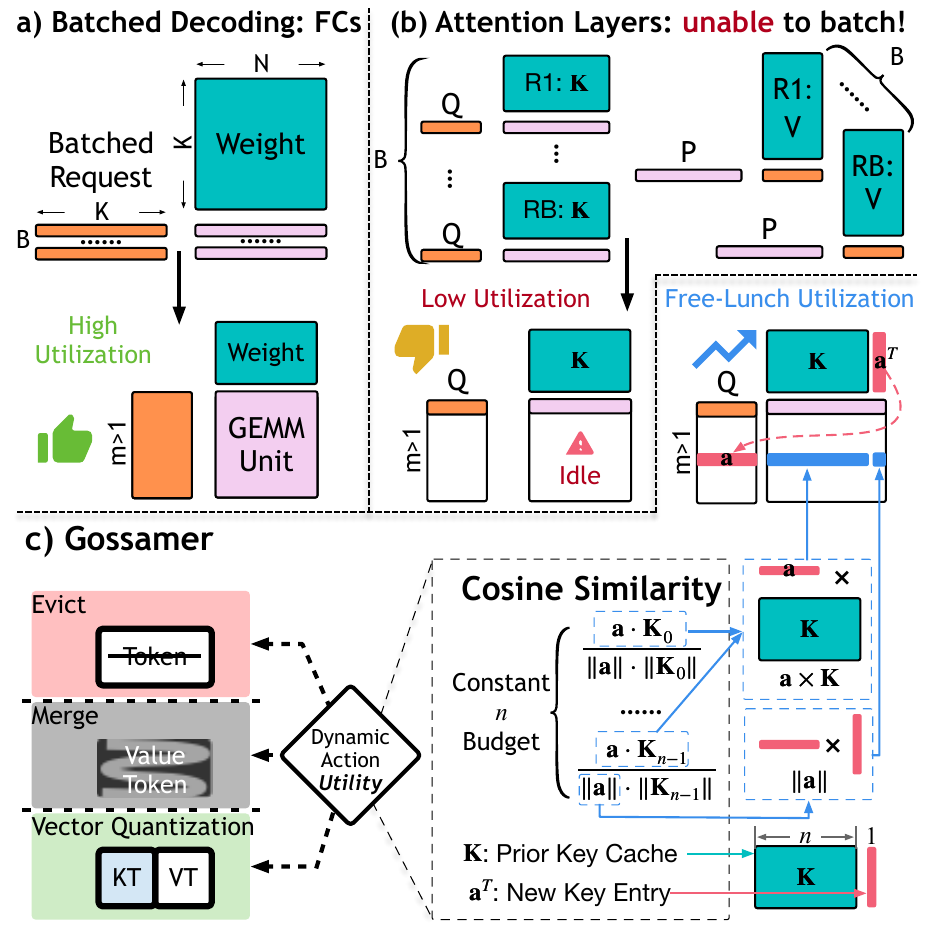}
    \caption{
  Overview of Gossamer: (a) batching boosts FC utilization, (b) attention idles enable {free-lunch utilization}, and (c) utility actions manage KV under a constant budget.
  }
  \vspace{-1.6em}
  \label{fig:intro}
\end{figure}

In this study, we propose \methodnamebold, a utility-driven KV cache architecture for reasoning LLMs that bridges the gap among algorithm, system, and hardware design.  
The utility formulation builds on comprehensive cross-domain observations, integrating insights from all three domains in a mutually reinforcing way.  

From a system perspective, current serving stacks rely on batching to improve inference throughput for reasoning workloads.  
As shown in \cref{fig:intro}-a, batching allows multiple requests to share GEMM operations in fully connected layers, yielding high utilization.
However, in attention layers, as shown in \cref{fig:intro}-b, batching is infeasible because each request maintains a private KV cache.  
The computation thus degenerates into a series of memory-bound GEMV operations, leaving substantial compute units idle. 
{As illustrated by the roofline model in \cref{fig:2_1_background14}-a, this traps the attention operation in the \textit{memory-bound} region, preventing it from reaching the hardware's theoretical compute roof.}

On the algorithmic side, Gossamer’s utility-based action framework unifies advanced KV cache compression techniques, including quantization, eviction, and merging.  
All these actions depend on cosine similarity between new and historical KV entries to identify redundancy and decide token-level operations.  
Ideally, each new token would perform such checks for optimal compression, but comparing against all cached entries is \textit{computationally expensive}, forcing existing systems to group tokens and delay updates.  

By combining architectural and algorithmic insights, we make a key observation.  
Cosine similarity can be decomposed into per-key computations that align naturally with attention operations.  
As illustrated in \cref{fig:intro}-c, the similarity between the new key $\mathbf{a}$ and historical keys $\mathbf{K}$ can be fused into the existing attention matrix multiplication, sharing intermediate products and compute flow.  
This enables \textit{cosine similarity calculation to be executed entirely within the attention kernel without additional latency or bandwidth cost}.  
{Crucially, because the baseline attention kernel is severely memory-bound, these fused operations remain well below the hardware's compute roof, meaning the extra computation comes essentially for free.}

We call this integration \textbf{Free-Lunch Utilization}.  
By exploiting idle compute cycles within the attention pipeline, \methodname performs similarity computation “for free,” maintaining global context visibility while hiding its cost.  
The resulting architecture supports dynamic utility-based actions such as quantization and eviction in place, with minimal overhead, and near-constant memory usage.  
Through this cross-domain co-design, \methodname achieves \textbf{global reasoning awareness with the efficiency of local in-place execution}, improving utilization across algorithm, system, and architecture levels.  

\noindent We make the following contributions in this paper.
\setlist{nolistsep}
\begin{itemize}[leftmargin=1.2em]
    \item We identify a fundamental bottleneck in KV cache compression for reasoning LLMs: existing designs face a dilemma between \textit{accurate but computationally expensive} global compression and \textit{faster but less accurate} local compression, leaving a gap between compression quality and system efficiency.
    \item We present \methodnamebold, a utility-driven framework for constant-budget KV cache management that decides how cached states should be maintained and transformed according to their \textit{estimated utility} during reasoning.
    \item We design a hardware-aware implementation of this framework that \textit{fuses} utility estimation into attention computation and enables low-overhead, fine-grained KV cache management during streaming decoding.
    \item We demonstrate on long-context reasoning benchmarks that \methodnameslim achieves near-uncompressed accuracy under tight memory budgets while significantly improving inference efficiency, achieving up to 16.7\x speedup over FullKV, 5.5\x and 4.9\x over Oaken and Ecco, and 1.8\x over R-KV.
\end{itemize}

\section{Background and Motivation}
\label{sec:background}

\subsection{Reasoning LLM Inference}
\label{ssec:reasoning_inference}

\begin{figure}[t]
  \centering
  \includegraphics[width=\linewidth]{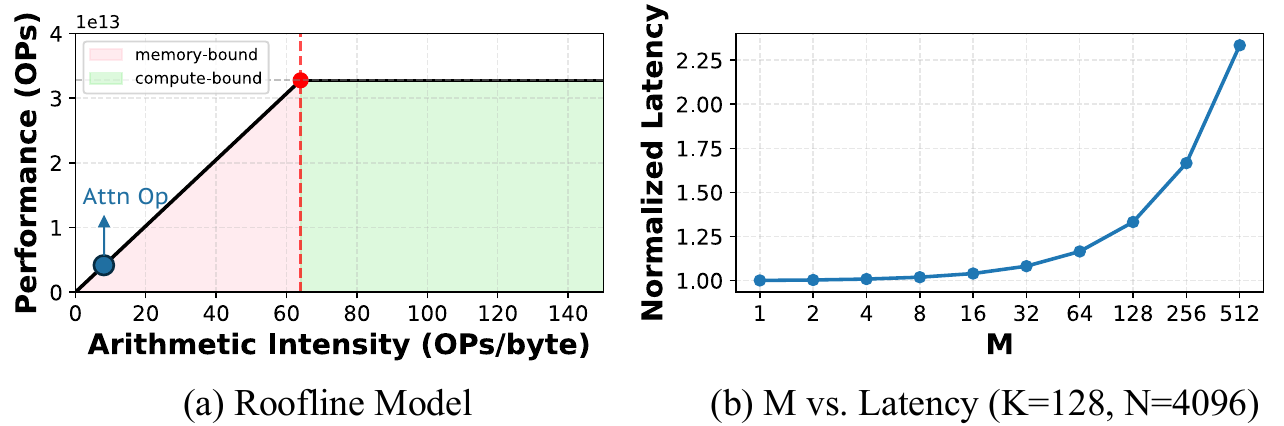}
  \caption{\textbf{(a)} Hardware Roofline model; \textbf{(b)} M vs. Latency}
  \vspace{-1.8em}
  \label{fig:2_1_background14}
\end{figure}

LLMs adopt the Transformer architecture~\cite{vaswani2017attention}.
Inference proceeds in two phases: \emph{prefill} and \emph{decode}~\cite{zhong2024distserve}. \textit{Prefill} processes the input prompt in parallel, while \textit{decode} generates one token at a time.
At decode step $T$, the query for the newly generated token is a row vector $Q_T\!\in\!\mathbb{R}^{1\times d_k}$, and the cached keys and values for that head are $K_T, V_T\!\in\!\mathbb{R}^{T\times d_k}$. 
Evaluating attention thus reduces to two matrix–vector products. 
As illustrated in \cref{fig:2_1_background14}-a, in GEMM terms, the left operand has an inherent $M{=}1$, which yields low arithmetic intensity and makes decode-time attention \emph{memory bound} rather than compute bound.

Reasoning-oriented models typically have a considerable generation length $T$, which has two critical system-level consequences. 
\emph{(i) HBM capacity and batching:} systems must preallocate KV cache space for the target generation length; because the cache grows linearly with $T$ per layer and head, HBM capacity quickly becomes the limiting resource~\cite{wang2024sofa,hooper2024kvquant}, constraining the maximum concurrent batch size and thus throttling throughput. 
\emph{(ii) Bandwidth at attention:} at each decode step, the accelerator must stream the cached $K, V$ tensors for all previous tokens to compute attention for the next token. Hence, the cost of attention grows with the effective KV length~\cite{pope2023efficiently}. 
Reducing this length directly lowers the number of bytes moved per generated token and speeds up memory-bound attention. 
Consequently, KV cache compression addresses both constraints simultaneously: it increases the batch size that fits in HBM and accelerates decode-time attention by shrinking the per-step KV footprint, reducing end-to-end latency with minimal impact on quality.

\noindent \textbf{Takeaway 1:} Decode-time attention is memory-bound; since the KV cache grows with sequence length, compressing it increases HBM-fit batch size and accelerates attention.

\subsection{KV Cache Compression}
\label{ssec:kv_compression}

Existing approaches to compress the KV cache fall into three broad categories. 
\emph{Eviction} methods~\cite{fu2024lazyllm,long2025sliminfer,zhang2023h2o} reduce the effective cache length by discarding tokens deemed less informative (e.g., oldest-first, stride-based thinning, entropy- or attention-driven policies, or head-specific retention).
\emph{Quantization} methods~\cite{lin2024qserve,hooper2024kvquant,zhang2405kv,cheng2025ecco,bwq,duan2026eva,olive,ant} shrink per-token storage by representing $K$ and $V$ at lower precision (e.g., per-head or group-wise scaling in 8/6/4-bit or FP8 formats, sometimes with outlier handling or error compensation) to cut both capacity and bandwidth. 
\emph{Merging} methods~\cite{wang2024model,zhang2024cam} aggregate multiple tokens into a single surrogate memory entry (e.g., temporal pooling or clustering within segments or per head), directly lowering the number of cache entries that attention must read each step.

Each family faces inherent trade-offs. 
Eviction reduces sequence length but risks dropping long-range dependencies and often requires carefully tuned, context-dependent policies. 
Quantization preserves length but can introduce accuracy loss from outliers and scale mismatch across layers and heads. 
Merging improves bandwidth by shortening the cache but may blur token-specific information and accumulate bias over depth. 
Because we must simultaneously address \emph{both} axes, bytes per token (capacity) and tokens per step (bandwidth), and because sensitivity varies by layer, head, and input regime, relying on a single technique rarely yields strong compression at fixed quality.

\noindent \textbf{Takeaway 2:} Effective KV compression should be hybrid and adaptive, combining eviction, quantization, and merging to reduce capacity and bandwidth while maintaining model quality.

\subsection{Limits of Scoring-Based KV Compression}
\label{ssec:compression_limitation}

\begin{figure}[t]
  \centering
  \includegraphics[width=\linewidth]{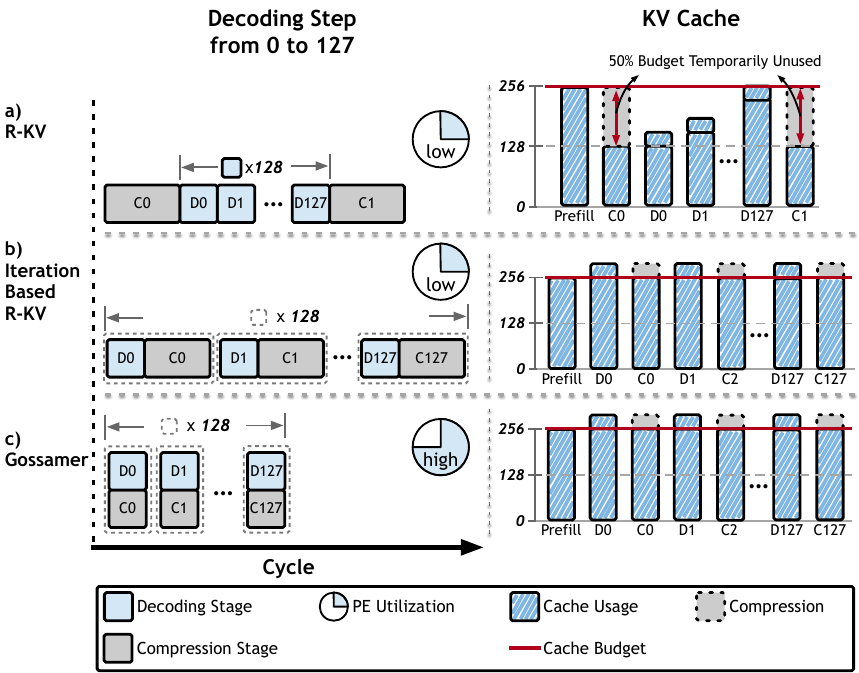}
  \caption{Limits of Scoring-Based KV Compression.}
  \vspace{-1.2em}
  \label{fig:2_2_background3}
\end{figure}

State-of-the-art KV compression often relies on per-token \emph{importance scoring} that ranks cache entries via global similarity or clustering. 
These schemes are not hardware friendly: score computation typically forms the all-pairs key similarity $S = KK^\top$ for $K \!\in\! \mathbb{R}^{T \times d_k}$, costing $\mathcal{O}(T^2 d_k)$ FLOPs plus extra memory traffic for reductions, top-$k$ selection, and index reshuffling. 
This overhead can easily exceed the per-step attention cost $\mathcal{O}(Td_k + Td_v)$. 
\cref{fig:2_2_background3} compares three approaches across decoding steps 0-127 under a 256-entry budget. 
The left panels show decoding/compression stages, with pie charts indicating PE utilization during attention computation.
The right panels track cache usage against the budget.

R-KV \cite{rkv} compresses lazily every 128 tokens, as shown in \cref{fig:2_2_background3}-a.
The cache fills to 256 entries, then compression abruptly drops it to 128 entries in one step.
This causes two problems: \textit{(i)} substantial context is discarded at once, degrading accuracy; \textit{(ii)} 50\% of the budget sits unused afterward, wasting memory capacity.
PE utilization during attention is low because compression runs as a separate, expensive phase.
Iteration based R-KV (\cref{fig:2_2_background3}-b) addresses the waste by compressing every step using the same global scoring.
While cache usage stays near the budget limit, this approach incurs prohibitive overhead: each step requires a full $\mathcal{O}(T^2 d_k)$ similarity pass in a separate kernel.
PE utilization during attention remains low because compression and attention remain decoupled.
The expensive scoring dominates runtime, making per-step compression impractical.

Our method achieves the best of both worlds through input fusion in \cref{fig:2_2_background3}-c.
By embedding lightweight utility estimation directly into the attention GEMM, it compresses every step with negligible overhead.
Cache usage stays near the budget limit (avoiding waste), compression decisions preserve informative tokens (maintaining accuracy), and PE utilization during attention is high because no separate compression phase is needed.
\methodname eliminates the tradeoff between compression frequency and overhead.

\noindent \textbf{Takeaway 3:} Lazy, scoring-heavy designs drop too much context when triggered, while per-step global scoring is prohibitively expensive, motivating streaming, hardware-friendly KV compression with negligible per-step cost.

\subsection{Leveraging Underutilized Compute in Decode-Time Attention}
\label{ssec:leverage_idle}

\begin{figure}[t]
  \centering
  \includegraphics[width=\linewidth]{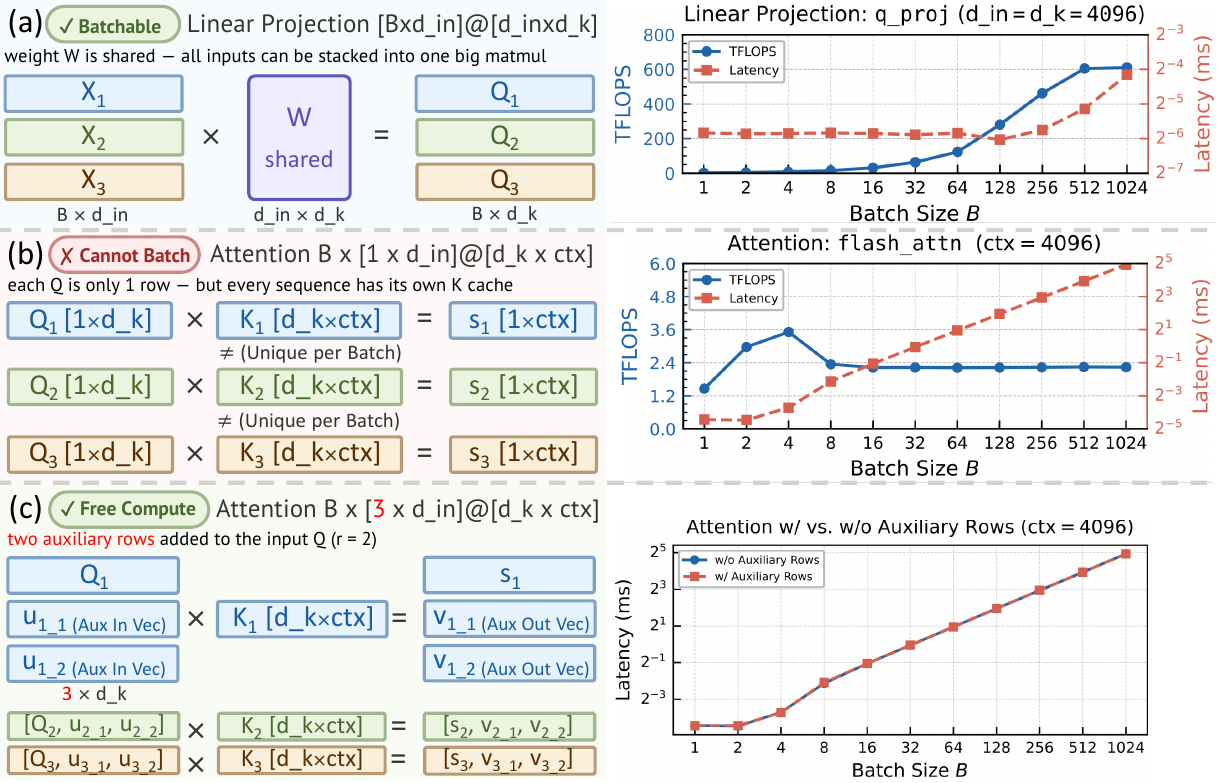}
  \caption{\Rebuttal{(a) Projection layers scale efficiently with batch size, achieving high TFLOPS utilization. (b) Attention layers cannot be batched across requests and remain memory-bound, with TFLOPS saturating regardless of batch size. (c) Adding auxiliary rows to the attention input incurs negligible latency overhead, validating the free-lunch utilization principle.}}
  \vspace{-1.8em}
  \label{fig:batchability}
\end{figure}

\Rebuttal{
Modern serving stacks lean on batching to raise throughput, yet batching helps the
two GEMM types inside a decoder layer in fundamentally different ways. In the projection
(fully-connected) layers, the weight matrix is \emph{shared} across requests, so the
activations of $B$ concurrent requests stack into a single large matmul. As shown in
\cref{fig:batchability}-a for the query projection, this drives the operation into the
compute-bound regime: achieved throughput climbs steeply with $B$ and approaches the
hardware's TFLOPS roof. Attention behaves oppositely, because each request owns a
\emph{private} $K/V$ cache and therefore cannot share an operand with any other request.
As shown in \cref{fig:batchability}-b, even under a highly optimized FlashAttention
kernel, the per-step computation reduces to a batch of \emph{independent}
$(1{\times}d)\!\cdot\!(d{\times}L)$ matrix--vector products, and its achieved TFLOPS
saturate at a low value regardless of $B$ while only latency grows. Attention thus
remains pinned in the memory-bound region even with batched-decode kernels, which means
the idle compute \methodname exploits is a \emph{structural} property of decode-time
attention rather than an artifact of single-request inference.
}

During decode, attention reduces to $(1{\times}d)\!\cdot\!(d{\times}L)$ matrix--vector products: with one query per step, the left operand has $M{=}1$, so compute units are underutilized while performance is dominated by reading the $L$ cached keys and values from memory. 
In this memory-bound regime, modest increases in M incur negligible latency cost, as demonstrated in \cref{fig:2_1_background14}-b. 

We exploit this by \emph{padding} the attention’s left operand with a few auxiliary vectors along the $M$ dimension. 
Let $K\!\in\!\mathbb{R}^{L\times d}$ be the cached keys and $q_t\!\in\!\mathbb{R}^{d}$ the current query.
Instead of computing only $q_t^\top K^\top/\sqrt{d}$, we stack $r$ lightweight vectors $u_1,\ldots,u_r\!\in\!\mathbb{R}^{d}$ (e.g., derived from the new token or simple running summaries) and form
\begin{equation}
Y_t \;=\;
\underbrace{\begin{bmatrix}
q_t^\top/\sqrt{d}\\[2pt]
u_1^\top\\
\vdots\\
u_r^\top
\end{bmatrix}}_{\displaystyle M_t \in \mathbb{R}^{(1+r)\times d}}
K^\top
\;\in\;\mathbb{R}^{(1+r)\times L}.
\end{equation}
The first row produces the usual attention pre-logits, while the additional rows yield per-position signals (e.g., similarity or redundancy proxies) that can be turned into streaming scores. 
Because $K^\top$ is read once regardless and $r$ is small, this fused multiply adds negligible overhead (no extra cache passes and only a small increase in FLOPs) while converting otherwise idle compute into fresh, per-step metrics. 
\Rebuttal{
In \methodname this slack is filled with just $r{=}2$ auxiliary rows, the new
normalized key and a running key accumulator, yielding the compact 3-row fused GEMM
detailed in \Cref{ssec:hw_fused}.
}

\Rebuttal{
Crucially, this argument survives batching. Since attention cannot be batched across
requests, each request independently keeps a tiny $M{=}1{+}r$ no matter how many requests
are served concurrently, so the padded rows \emph{never} push attention out of its memory-bound
operating point. \Cref{fig:batchability}-c confirms this empirically: across the full
batch-size sweep, decode latency with and without the $r$ auxiliary rows is virtually
indistinguishable, demonstrating that the fused similarity computation is obtained
essentially for free under realistic serving conditions, not merely for a single
request.

}

\noindent \textbf{Takeaway 4:} 
Decode attention with $M{=}1$ leaves compute resources underutilized; adding a few auxiliary rows ($M{\rightarrow}1{+}r$) utilizes this slack to generate per-step similarity scores, realizing ``free-lunch'' utilization. \Rebuttal{Because attention is non-batchable across requests, this slack (and hence the
free-lunch opportunity) persists even under high-throughput batched serving.}

\section{Gossamer: Compression Methodology}
\label{sec:methodology}

Our KV cache management system is a multi-stage, utility-driven framework designed to maximize memory efficiency while preserving model performance. 
The framework first identifies a core set of high-value tokens for full-precision retention using the token prioritization algorithm detailed in \Cref{ssec:scoring}. 
All other tokens are then processed by a dynamic action utility model (\Cref{ssec:dau_framework}) that selects an appropriate compression action under a fixed KV budget.
This modular design allows us to apply compression techniques, such as the vector quantization and merging described in \Cref{ssec:vq} and \Cref{ssec:merging}, only to tokens that are poor candidates for full retention but are still too valuable to evict.

\subsection{Candidate Selection: Scoring Foundation}
\label{ssec:scoring}

We base our candidate selection on prior KV scoring schemes driven by importance and redundancy~\cite{rkv, li2024snapkv, pyramidkv}.
\Cref{alg:token_scoring} assigns per-head token priorities by balancing \emph{importance} against \emph{redundancy}. 

\emph{Importance} is computed as the mean attention each token receives from a fixed window of the last \Rebuttal{$\tau=8$} queries, yielding a length-$L$ vector that summarizes how much recent decoding has relied on each cached token.

\emph{Redundancy} is estimated using cosine-similarity-based signals over the key vectors: for each token, we measure its similarity to the rest of the cache and aggregate these values into an average redundancy score, which is then normalized via a softmax to make scores comparable across tokens.
Higher redundancy indicates greater semantic overlap with other entries.

\emph{Priority} combines these two signals using a trade-off parameter $\beta\!\in\![0,1]$, increasing with importance and decreasing with redundancy.
Tokens are ranked by the resulting priority scores, and the top $B_{\text{main}}$ entries are retained in the main full-precision KV cache $C_{\text{main}}$, while the remaining tokens become candidates for compressed storage or eviction in the later stages of \methodname.

\begin{algorithm}[t]
\caption{Token Prioritization Scoring (per head $h$)}
\label{alg:token_scoring}
\begin{algorithmic}[1]
\Input $K^h \in \mathbb{R}^{L\times d}$ \Comment{Key vectors}
\Input $\mathcal{A}^h \in \mathbb{R}^{\tau\times L}$ \Comment{Attention from last $\tau$ queries}
\Input $\beta \in [0,1]$ \Comment{Importance–redundancy trade-off}
\Output $\Psi^h \in \mathbb{R}^{L}$ \Comment{Priority scores}

\State \textbf{Phase 1: Importance}
\State $\omega^h \gets \tfrac{1}{\tau}\,\mathbf{1}_\tau^\top \mathcal{A}^h$ \Comment{Mean attention per token}

\State \textbf{Phase 2: Redundancy}
\State $\Sigma^h \gets \Call{CosineSim}{K^h, K^h}$ \Comment{$L\times L$ similarity}
\State \Call{ZeroDiagonal}{$\Sigma^h$} \Comment{Remove self-similarity}
\State $\bar{\sigma}^h \gets \tfrac{1}{L-1}\,\Sigma^h \mathbf{1}_L$ \Comment{Row-wise totals}
\State $\rho^h \gets \Call{softmax}{\bar{\sigma}^h}$ \Comment{Normalize redundancy}

\State \textbf{Phase 3: Final scores}
\State $\Psi^h \gets \beta\,\omega^h - (1-\beta)\,\rho^h$
\State \Return $\Psi^h$
\end{algorithmic}
\end{algorithm}

\Cref{alg:token_scoring} describes the ideal importance-redundancy scoring we aim to compute.
Naively, this requires $\mathcal{O}(L^2 d_k)$  similarity computation.
In our hardware implementation (Section IV), we avoid materializing $\Sigma^h$ by using normalized keys, a running key accumulator, and per-token top-k similarity tables.
This allows importance and redundancy scores to be maintained from a single fused attention GEMM, eliminating extra passes over the KV cache.

\subsection{Compression Action 1: Vector Quantization}
\label{ssec:vq}

To reduce the significant memory footprint of the KV cache, we employ a novel vector quantization (VQ) strategy that combines \textbf{correlation-based channel reordering} with \textbf{gain-shape quantization}. 
Our objective is to compress the per-head $D$-dimensional KV vectors with high fidelity while preserving simple, regular access patterns for hardware.

\subsubsection{Sub-vector Quantization}
\label{ssec:sub_vq}

Directly applying VQ to a high-dimensional vector $x \in \mathbb{R}^D$ (e.g., $D=128$) is statistically and computationally infeasible due to the curse of dimensionality. 
We adopt sub-vector quantization, partitioning x into M non-overlapping sub-vectors of dimension G:
\[
x = [x^{(1)}, x^{(2)}, \dots, x^{(M)}], \text{where } x^{(i)} \in \mathbb{R}^G \text{ and } D = M \times G
\]
We use G=8, a D=128 vector splits into M=16 sub-vectors, each with a separate codebook $\mathcal{C}_i$.

\subsubsection{Correlation-Based Reordering}
\label{ssec:reordering}

Naive sub-vector quantization suffers from arbitrary partitioning: consecutive channels (e.g., dimensions 0-7, 8-15) are not necessarily correlated, leading to suboptimal quantization. 
\cref{fig:3_1_vq}-b (left) shows VQ on adjacent channels 49 and 50, yielding an average error of 0.118.
\cref{fig:3_1_vq}-b (right) shows VQ on correlated channels 46 and 50, reducing error to 0.035, which is a 3.4\x improvement. 
Grouping correlated dimensions creates tighter clusters and significantly improves quantization quality.

We address this through offline correlation-based reordering. 
For each attention head, we compute the $D \times D$ channel correlation matrix $R$ from profiled KV cache data, then apply greedy clustering to find a permutation $\pi$ grouping the $G$ most correlated channels.
The algorithm iteratively builds size-$G$ groups by selecting the most correlated unassigned pair, then greedily adding dimensions with the highest average correlation to existing members.

Permutations are applied \textit{offline} to eliminate runtime overhead. 
We derive $\pi_K$ from K-cache correlations and $\pi_V$ from V-cache correlations, then pre-emptively reorder the columns of projection matrices $W_Q$, $W_K$, and $W_V$ accordingly. 
At inference, produced vectors $q_{\pi_K}$, $k_{\pi_K}$, and $v_{\pi_V}$ are already optimally ordered and directly partitioned into sequential sub-vectors for quantization. 
This ensures each sub-vector contains highly correlated dimensions by construction, reducing quantization error as shown in \cref{fig:3_1_vq}.

\subsubsection{Gain-Shape Codebook Generation}
\label{ssec:gain-shape}

\begin{figure}[t]
  \centering
  \includegraphics[width=\linewidth]{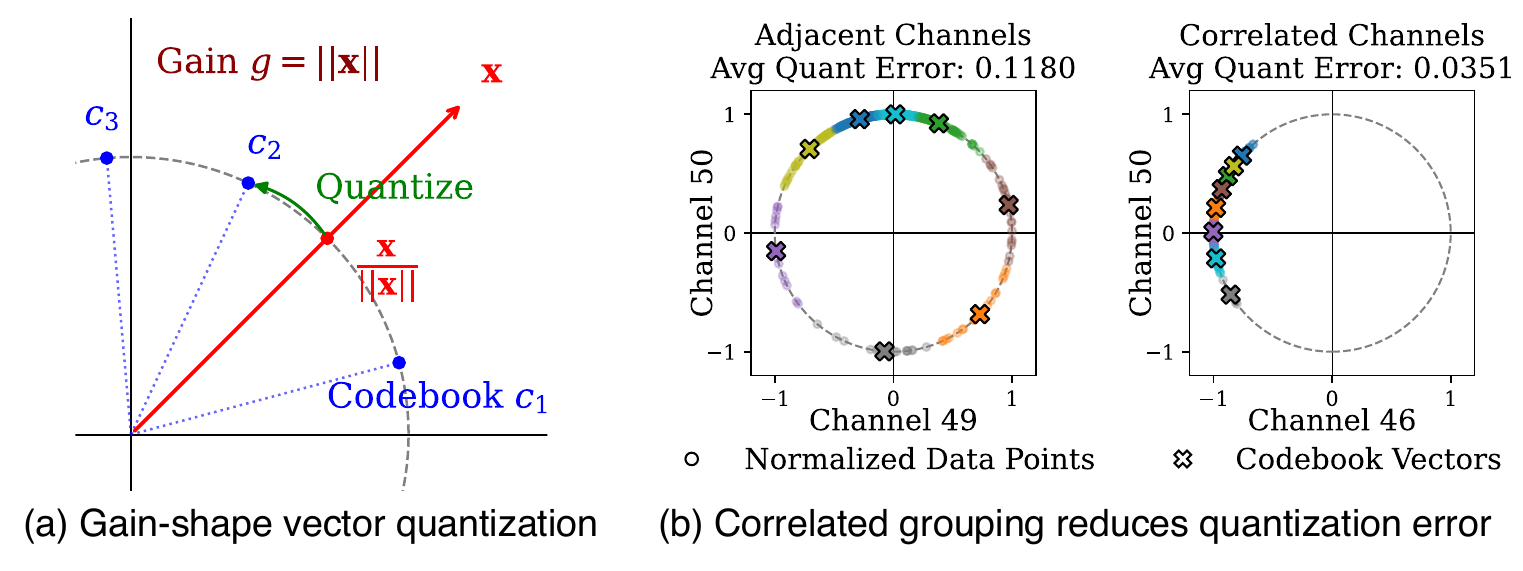}
  \caption{Gain-shape vector quantization and correlation-based channel reordering.}
  % \vspace{-2em}
  \label{fig:3_1_vq}
\end{figure}

We further improve quantization accuracy by using a gain-shape formulation, as illustrated in \cref{fig:3_1_vq}-a. 
Instead of quantizing the (permuted) sub-vector $x_{\pi}^{(i)}$ directly, we decouple its magnitude (gain) from its direction (shape).
The gain $g^{(i)}$ is defined as the $L_2$ norm of the sub-vector, and the shape $\hat{x}_{\pi}^{(i)}$ is the corresponding unit-norm vector:
\begin{align*}
    g^{(i)} = \|x_{\pi}^{(i)}\|_2 \qquad \hat{x}_{\pi}^{(i)} = \frac{x_{\pi}^{(i)}}{g^{(i)} + \epsilon}
\end{align*}
where $\epsilon$ is a small constant (e.g., $10^{-12}$) for numerical stability.

Our offline codebook generation process operates on these shape vectors. 
For each sub-vector position $i \in [1, M]$, we train a dedicated codebook $\mathcal{C}_i \in \mathbb{R}^{K \times G}$ by applying the $k$-means algorithm to the set of all shape vectors $\{\hat{x}_{\pi}^{(i)}\}$ from the profiled KV cache.
We use $K=256$ clusters, allowing each shape vector to be represented by an 8-bit index.

\subsubsection{Online Encoding and Storage}
\label{ssec:online-encoding}
During inference, when the dynamic action--utility (DAU) framework selects a candidate token $t_c$ for quantization, its key vector $k_c$ (and similarly its value vector $v_c$) undergoes an encoding process. 
Because of the offline matrix reordering (\Cref{ssec:reordering}), $k_c$ is already the permuted vector $k_{\pi_K}$. 
We first partition $k_c$ into $M = 16$ sub-vectors $\{k_{c,1}, \dots, k_{c,16}\}$, each of dimension $G = 8$.
For each sub-vector $k_{c,i}$, we compute its gain $g_i = \|k_{c,i}\|_2$.
Since codebook entries $c_{i,j} \in \mathcal{C}_i$ are unit-norm, selecting the best codeword by cosine similarity reduces to maximizing the dot product:
\[
s_i = \max_j k_{c,i}^\top c_{i,j},
\]
We store the corresponding index $j^\star$ as the 8-bit code for that sub-vector, and store it together with the gain $g_i$.

To inform the DAU framework, we derive a quantization fidelity score:
\[
\phi_{\text{quant}}(k_c)
=
\frac{1}{\sqrt{M}\,\|k_c\|_2}
\sum_{i=1}^{M} s_i
\]
which aggregates the best dot products across sub-vectors, normalized by the full vector norm and $\sqrt{M}$ for consistent scaling across sub-vectors.

The quantized representation consists of $M=16$ FP16 gains (32 bytes) and $M=16$ 8-bit indices (16 bytes), totaling 48 bytes versus 256 bytes for the original FP16 vector, achieving 5.3\x compression.
The compressed token moves from main cache $C_{\text{main}}$ to quantized cache $C_{\text{quant}}$; the same procedure applies to $v_c$.

\subsection{Compression Action 2: Merging}
\label{ssec:merging}
This action aims to preserve the information of the candidate token by consolidating it with a semantically similar token that is being retained in the cache.
The operational details of this process are as follows:

\subsubsection{Target Search}
When a candidate token $t_c$ (with key vector $k_c$ and value vector $v_c$) is considered for merging, the system first identifies the most suitable merge target, $t_t$, from the set of tokens currently in the main retained cache, $C_{main}$.
The target is found by performing a nearest-neighbor search in the key-vector space using cosine similarity as the metric:
$$
t_t = \arg\max_{t_i \in C_\text{main}} \text{sim}(k_c, k_i)
$$

\subsubsection{Similarity Score Calculation}
The maximum similarity value found in the search is recorded as the merge similarity score for this action, which is used by the DAU framework to evaluate this choice:
$$
\phi_\text{merge}(k_c, k_t) = \text{sim}(k_c, k_t) = \frac{k_c \cdot k_t}{\|k_c\| \cdot \|k_t\|}
$$

\subsubsection{State Update}
If the merge action is selected, the state of the target token $t_t$ is updated to incorporate the information from the candidate $t_c$. 
Consistent with the goal of preserving the target's semantic anchor in the cache, the key vector $k_t$ remains unchanged. 

The value vector $v_t$, however, is updated to a new representation, $v_t^{\text{new}}$. 
This process merges the two vectors while preserving the $L_2$ norm of the original target token. 
First, an intermediate vector is computed by summing the source and target value vectors $v_c$ and $v_t$.
This vector is then normalized and rescaled to match the original magnitude of $v_t$:
$$
v_t^{\text{new}} = \|v_t\|_2 \cdot \frac{v_t + v_c}{\|v_t + v_c\|_2 + \epsilon}
$$
where $\epsilon$ is a small constant for numerical stability. 
This operation consolidates the information from $t_c$ into $t_t$ while maintaining the original target's norm, after which the candidate token $t_c$ is evicted from the cache.

\subsection{The Dynamic Action-Utility Framework}
\label{ssec:dau_framework}

Tokens that are \textit{not} selected for retention by \Cref{alg:token_scoring} (i.e., those with the lowest priority scores $\Psi^h$) are not immediately discarded.
Instead, they become candidates for a set of memory-saving actions.
The central innovation of this work is the DAU decision framework, which provides a principled logic for selecting the most appropriate action by transforming the choice into a formal cost-benefit analysis.
\Rebuttal{Since no evicted or compressed token can later be recovered, the DAU is designed to minimize the information loss from each irreversible choice.}

\subsubsection{Cost-Benefit Utility Formulation}
\label{sssec:cost_benefit}

The DAU framework models KV cache management as a resource allocation problem.
When the main cache $C_{main}$ exceeds its budget, the candidate token $t_c$ (identified as the one with the lowest priority score $\Psi(t_c)$) must be acted upon.
The framework formalizes the decision as choosing an action $a$ from the set $A = \{\text{Evict, Merge, Quantize}\}$ that maximizes a utility function $U(a, t_c)$.
\begin{equation}
\label{eq:utility_fn_base}
a^* = \underset{a \in A}{\arg\max} U(a, t_c)
\end{equation}
The utility function reflects the net benefit of an action.
An ideal, lossless action (e.g., retaining $t_c$ in full precision) would have a utility of zero.
All our defined actions incur some information loss and thus have a \textit{non-positive utility} (i.e., $U \le 0$).
The magnitude of the utility corresponds to the severity of the loss. This allows us to frame the decision as choosing the ``least bad" action, the one that minimizes the damage to the model's future predictive capabilities.

\subsubsection{Action Utility Definitions}
\label{sssec:action_utilities}

The utilities for Evict and Merge are straightforward.
The Quantize utility is more complex as it must account for a secondary, constrained budget.

\noindent\textbf{a) Evict Utility ($U_{evict}$):}
Eviction results in the complete and irreversible loss of $t_c$. The cost of this action is therefore directly proportional to the importance of the token.
The utility is defined as the negative of its importance score, $\omega(t_c)$, which was computed in \Cref{alg:token_scoring}:
\begin{equation}
\label{eq:utility_evict}
U_\text{evict}(t_c) = -w_\text{imp} \cdot \omega(t_c)
\end{equation}
Here, $w_\text{imp}$ is a positive hyperparameter that scales the importance score into the utility space.

\noindent\textbf{b) Merge Utility ($U_\text{merge}$):}
Merging consolidates $t_c$ into a semantically similar target token $t_t$ (which is already in $C_{main}$).
We define $\phi_\text{merge}(t_c, t_t)$ as their cosine similarity.
The information loss is related to the dissimilarity, so the utility is defined by the deviation from a perfect merge:
\begin{equation}
\label{eq:utility_merge}
U_\text{merge}(t_c) = w_\text{merge} \cdot (\phi_\text{merge}(t_c, t_t) - 1)
\end{equation}
The term $(\phi_{merge} - 1)$ is always non-positive, yielding $U=0$ for a perfect merge.

\noindent\textbf{c) Adjusted Quantize Utility ($U'_\text{quantize}$):}
The Quantize action is unique in that it interacts with two separate budgets: it resolves the $C_\text{main}$ budget overflow but, in turn, adds an entry to the quantized cache, $C_\text{quant}$.
If $C_\text{quant}$ is also full, a token must be displaced from it.

We term the cost of this forced eviction the Displacement Cost ($D_\text{cost}$).
This is the opportunity cost of quantizing $t_c$, equal to the value of the token that must be removed from $C_\text{quant}$ to make room.
We evict the least valuable token, $t_{lq}$, currently in the quantized cache, where $t_{lq} = \arg\min_{t_i \in C_\text{quant}} \omega(t_i)$.
The cost is simply this token's eviction utility:
\begin{equation}
\label{eq:d_cost}
D_\text{cost} = U_\text{evict}(t_{lq}) = -w_\text{imp} \cdot \omega(t_{lq})
\end{equation}
This cost, which is always negative, is added to the base quantization utility, $U_\text{quantize}(t_c)$.
The base utility is calculated similarly to the merge utility, based on the fidelity $\phi_\text{quant}(t_c)$ (e.g., cosine similarity) between the original vector and its quantized reconstruction:
\begin{equation}
\label{eq:utility_quant_base}
U_\text{quantize}(t_c) = w_\text{quant} \cdot (\phi_\text{quant}(t_c) - 1)
\end{equation}
The final, \textit{adjusted} quantization utility $U'_\text{quantize}$ incorporates the $D_\text{cost}$ only if $C_\text{quant}$ is at its budget limit:
\begin{equation}
\label{eq:utility_quant_adj}
U'_\text{quantize}(t_c) = \begin{cases} 
      U_\text{quantize}(t_c) + D_\text{cost} & \text{if } |C_\text{quant}| \ge B_\text{quant} \\
      U_\text{quantize}(t_c) & \text{if } |C_\text{quant}| < B_\text{quant}
   \end{cases}
\end{equation}
This mechanism creates a dynamic, self-regulating system.
As $C_\text{quant}$ fills with more important tokens, $D_\text{cost}$ becomes a larger penalty, making quantization a less attractive option.

\subsubsection{Final DAU Decision Rule}
\label{sssec:final_rule}

By combining the action-specific utilities with the budget-aware displacement cost, we arrive at the final, comprehensive decision rule.
When the main cache $|C_\text{main}|$ exceeds its budget $B_\text{main}$, the candidate token $t_c = \arg\min_{t_i \in C_\text{main}} \Psi(t_i)$ is identified.
The framework then evaluates the final utility for all three actions and selects the one with the highest (least negative) utility:
\begin{equation}
\label{eq:utility_fn_final}
a^{*} = \arg\max_{a \in A} \{U_\text{evict}(t_c), U_\text{merge}(t_c), U'_\text{quantize}(t_c)\}
\end{equation}
This rule transforms compression into a clear optimization problem that manages contention across both cache budgets.
Critically, every action that resolves a $|C_\text{main}| > B_\text{main}$ overflow either deletes $t_c$ from $C_\text{main}$ or migrates it to $C_\text{quant}$.
Because $C_\text{quant}$ is itself capped at $B_\text{quant}$ with displacement enforced through the utility framework, the total number of KV entries stored across both caches is bounded by $B_\text{main}$ + $B_\text{quant}$ for all decode steps.
This ensures constant memory usage regardless of sequence length T, fulfilling our design goal of a predictable, bounded memory footprint for long-form reasoning workloads.

\section{Gossamer: Architecture Design}
\label{sec:hardware}

To support the dynamic, utility-driven KV cache management framework described in \Cref{sec:methodology}, we co-design a specialized hardware accelerator.
This accelerator integrates cache management logic directly into the attention datapath, minimizing latency and compute overhead.
The design centers on three key ideas:
\begin{itemize}
    \item[(i)] \textit{Decoupled key storage:} Storing $L_2$-normalized keys ($\tilde{k}$) and their norms ($\kappa$) separately, which \textbf{allows dot-products to serve as cosine similarities}.
    \item[(ii)] \textit{Lightweight metadata:} Maintaining lightweight metadata (e.g., a key accumulator $s^h$ and a tiny similarity table) that enables computation of importance and redundancy scores \textbf{without re-reading the full key cache}.
    \item[(iii)] \textit{Fused computation:} Fusing the computations for attention, redundancy, and neighbor search into a \textbf{single, efficient matrix multiplication}.
\end{itemize}

\begin{figure}[t]
  \centering
  \includegraphics[width=\linewidth]{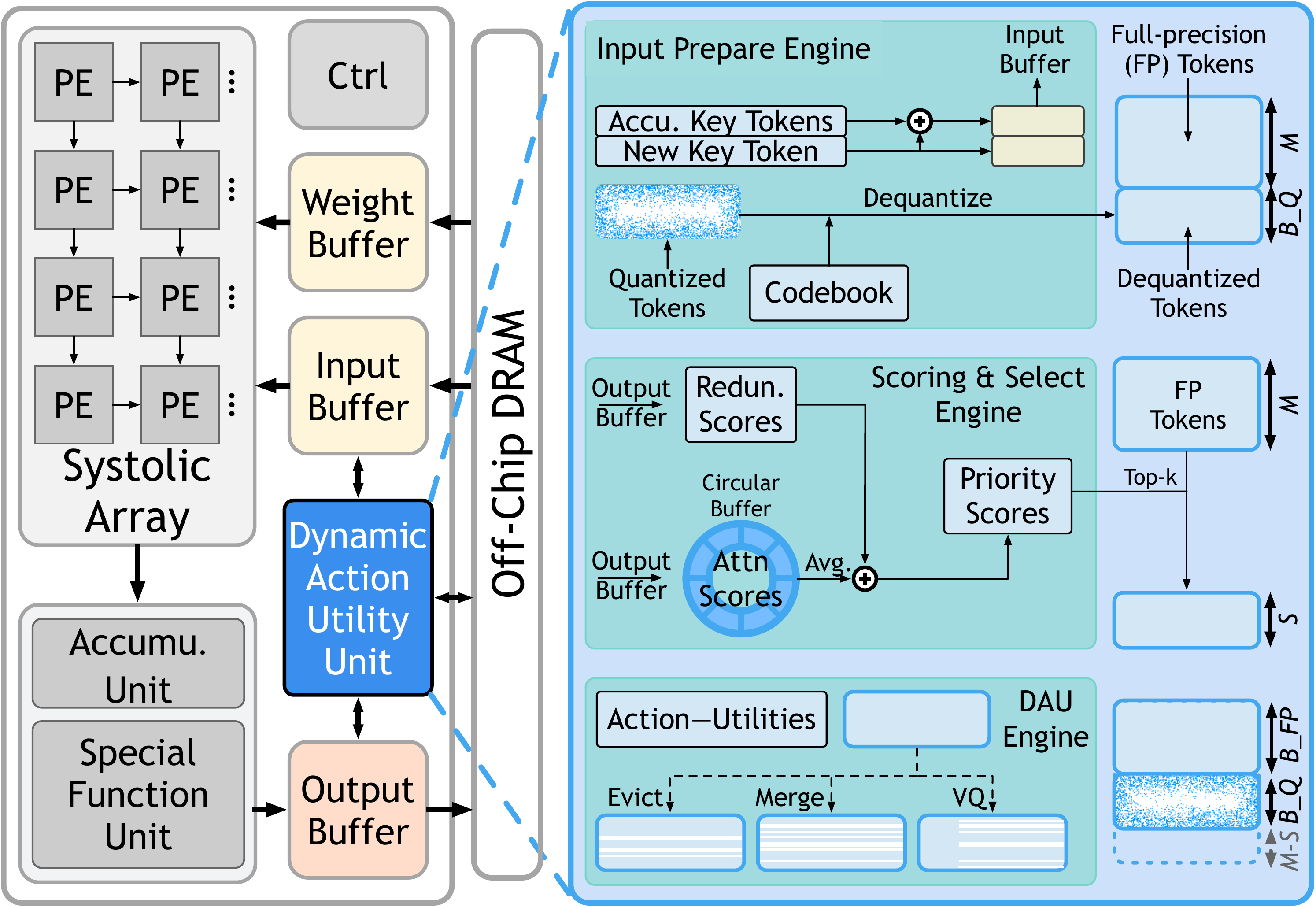}
  \caption{Overview of the Gossamer Architecture.}
  \label{fig:5_1_overview}
\end{figure}

\subsection{Gossamer Overview}
\label{ssec:gossamer_overview}

The \methodname architecture is wrapped around a conventional TPU-like systolic array~\cite{jouppi2017datacenter,kung1979systolic}, as shown in \cref{fig:5_1_overview}. This core features a standard matrix-multiply unit, complete with input, weight, and output buffers, which computes query-key products and value-weighted outputs.
\methodname augments this baseline design with three specialized components. 

First, an \textbf{input prepare engine} normalizes new keys, updates an accumulated key register, concatenates them with the new query, and writes the result into the input buffer; this engine also dequantizes tokens and feeds them into the weight buffer.
Second, a \textbf{scoring and select engine} receives redundancy and importance scores from the output buffer, updates the system metadata, and uses these scores to select low-priority tokens.
Finally, a \textbf{DAU engine} consumes this metadata, performing simple calculations to decide whether to evict, merge, or quantize each low-priority token.

\subsection{Hardware-Managed Metadata}
\label{ssec:hw_metadata}

To manage the cache efficiently, the hardware maintains several key metadata structures for each attention head.
First is the decoupled key cache.
Instead of storing the full-dimensional key, the hardware splits it into its $L_2$ norm and its normalized shape vector. 
The main key cache buffer, therefore, stores only these unit-norm shape vectors.
The hardware also maintains a top-k similarity cache.
This is a small, fast buffer that stores the identifying indices and similarity scores for the $k_{sim}$ most similar tokens of each entry (e.g., $k_{sim}=3$).

Additionally, a single, full-precision register, called the accumulated key register, keeps a running vector sum of all the normalized key vectors currently in the main cache.
This accumulator is what allows for the efficient redundancy calculation later.
Finally, an attention history buffer, operating as a circular buffer, stores the most recent attention probability vectors, which are used to calculate the rolling average for the token importance score.

\begin{figure*}[t]
  \centering
  \includegraphics[width=\linewidth]{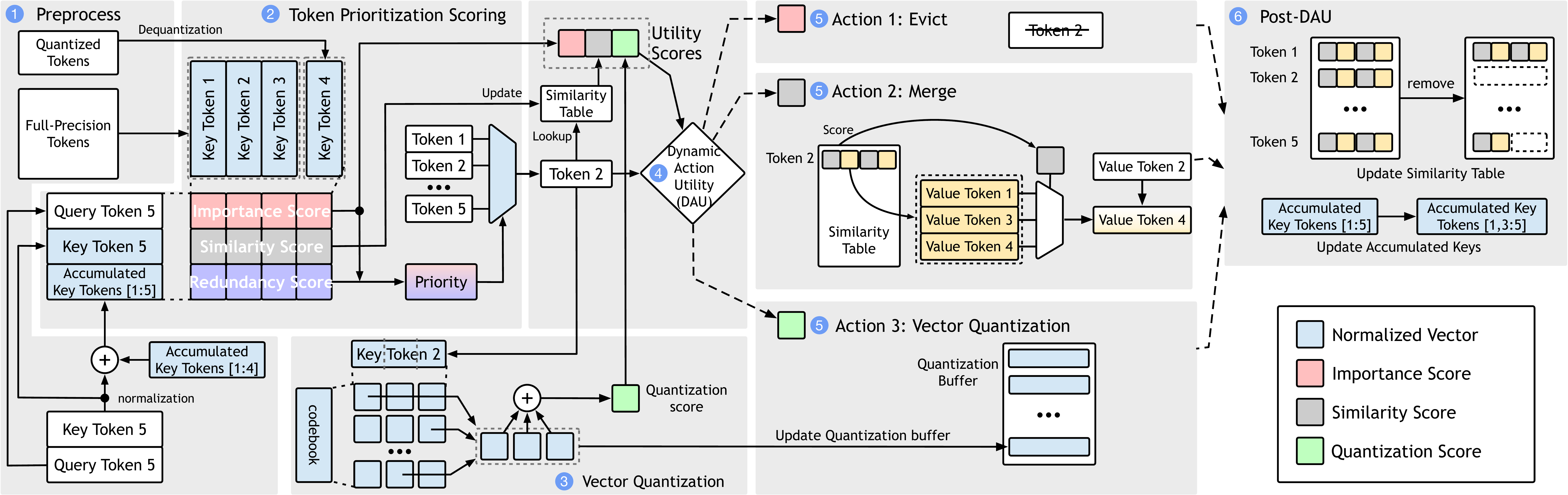}
  \caption{Detailed Architecture and Datapath of the Gossamer Framework.}
  \label{fig:5_2_datapath}
  \vspace{-1.2em}
\end{figure*}

\subsection{Pre-Attention Token Processing}
\label{ssec:hw_pre_attention}

When a new token arrives, it is first passed through the key projection matrix to generate a new key vector.
As shown in \cref{fig:5_2_datapath}-\bluecircle{1}, the hardware datapath intercepts this vector and immediately calculates its $L_2$ norm and its normalized shape vector.
This maintains the gain-shape decoupled representation.
The gain is appended to the metadata, and the shape vector is appended to the main key cache buffer.
At the same time, the accumulated key register is updated by adding this new normalized shape vector to its existing sum.
These steps prepare the three critical inputs for the next stage: \textit{the new query}, \textit{the new normalized key}, and \textit{the updated accumulated key vector}. 

These three vectors are then passed together into the input buffer for the fused computation.
During this process, the datapath also handles dequantization, loading tokens from the quantized cache and converting them back to full precision before they are used.
It is important to note that no runtime permutation overhead is required.
The correlation-based reordering of vector sub-parts is applied offline by pre-permuting the columns of the main query, key, and value projection matrices.
This means the query, key, and value vectors arrive from the projection layers already in their optimally permuted order, ready for direct use.

\subsection{Fused Attention and Score Computation}
\label{ssec:hw_fused}

As shown in \cref{fig:5_2_datapath}-\bluecircle{2}, our fused computation takes three inputs: the current query, the new key token, and the accumulated key vector (which represents the sum of all keys currently in the cache).
These three vectors are bundled into a small 3-row matrix and then multiplied against the entire matrix of normalized keys from the KV cache.
This single, efficient matrix multiplication simultaneously produces three distinct and vital results in its three output rows, all in one pass.

The first output row is used for the standard attention calculation.
It contains the initial query-key products.
Because the keys were normalized, we recover the true attention logits by scaling them with their corresponding L2 norm values, which are stored separately.
These true logits are then passed to a softmax function to produce the final attention scores.
In parallel, these scores are also used to update the importance score for each token, which is maintained as a streaming average of its recent attention values.

The second output row contains the cosine similarities between the new key and all other keys already in the cache.
This row is used for similarity tracking.
A top-k unit processes this data to find the most similar neighbors for the new token, populating its neighbor table.
Concurrently, this new similarity value is compared against the tables of all existing tokens, allowing their neighbor summaries to be kept fresh without requiring a full, costly rescan of the cache.

The third output row provides a highly efficient way to calculate redundancy by computing the dot product of the accumulated key vector with every key in the cache.
This serves as a streaming proxy for the row-sum of $\Sigma^h$ in \Cref{alg:token_scoring}: a token's total redundancy (its similarity to all other tokens) can be computed by taking this dot product and subtracting 1, completely avoiding the $\mathcal{O}(L^2 d_k)$ cost of building a full token-to-token similarity matrix.
The hardware processes these values into an average redundancy, then applies softmax to generate the final redundancy score.

Finally, the importance and redundancy scores are weighted and added together to obtain a final priority score for each token.
A comparator unit identifies the token with the lowest priority, which is then fed to the DAU engine for a final decision on whether to evict, merge, or quantize it.

\subsection{DAU-Driven Cache Management}
\label{ssec:hw_dau}

When the main cache budget is exceeded, the DAU engine activates.
As shown in \cref{fig:5_2_datapath}-\bluecircle{4}, following candidate identification, the hardware concurrently evaluates the utility of all three potential actions, evict, merge, and quantize, in parallel.
The evict utility is the most direct, computed from the candidate token's importance score.
The merge utility is determined by first retrieving the required merge similarity.
This value is obtained instantaneously by reading the maximum similarity score from the candidate token's pre-computed top-k similarity metadata table, which is then used to calculate the final utility of a merge operation.

Finally, the quantization utility (\cref{fig:5_2_datapath}-\bluecircle{3}) is assessed by the quantization engine, which computes a quantization fidelity score for the candidate token.
This fidelity score is obtained efficiently with no extra cost as a natural byproduct of the codebook lookup process itself, which inherently maximizes cosine similarity.
The resulting utility score is then adjusted to account for a potential displacement cost, which is only incurred if the dedicated quantization cache is already full and an existing quantized token must be evicted to make space.

A comparator unit then selects the single action that yields the highest utility, as illustrated in \cref{fig:5_2_datapath}-\bluecircle{5}.
If the selected action is to merge, the hardware reads the corresponding target token, executes the norm-preserving value vector update, and writes the updated value back to the main cache.
If the selected action is to quantize, the candidate token is encoded and moved to the quantization cache.

\subsection{Post-Action Maintenance}
\label{ssec:hw_post}

As demonstrated in \cref{fig:5_2_datapath}-\bluecircle{6}, regardless of the action taken (evict, merge, or quantize), the hardware performs a consistent set of metadata updates.
First, the accumulated key vector is corrected by subtracting the normalized key of the candidate token that was just processed.
This step is critical as it ensures that all subsequent redundancy calculations, which rely on this accumulator, remain exact.

Second, the system updates the neighbor tables.
Any table entries that referenced the processed token are invalidated by zeroing their scores and setting an invalid bit. 
Refreshing these entries immediately would be costly; instead, they are updated opportunistically in a later iteration. 
The attention buffer also slides forward automatically as part of this standard update.

\section{Evaluation}
\label{sec:evaluation}

\subsection{Accuracy Evaluation}

\textbf{Models and Datasets.} We evaluate using two distilled reasoning language models, {DeepSeek-R1-Distill-Llama-8B~\cite{guo2025deepseekr1distillllama8b}} and {DeepSeek-R1-Distill-Qwen-14B~\cite{deepseek2025r1distillqwen14b}}, on the {MATH-500~\cite{huggingface2025math500}} and {AIME 2024~\cite{maa2024aime}} benchmarks.
All checkpoints are obtained via HuggingFace Transformers~\cite{wolf2020transformers}. 
We further profile each model’s KV cache on the GSM8K~\cite{cobbe2021training} dataset and use the resulting statistics to build the quantization codebook.
For decoding, we cap the maximum generation length at 8,192 tokens for MATH-500 and 16,384 tokens for AIME 2024; increases beyond these limits did not improve accuracy in our preliminary sweeps.
Because greedy decoding can yield high variance for long-form reasoning models, and following prior work~\cite{deepseek2025deepseek_r1,rkv}, we adopt a pass@k protocol~\cite{li2022evaluating} and report pass@1 under stochastic sampling.
For the DeepSeek-R1 distilled checkpoints, we use the recommended hyperparameters: temperature $=0.6$ and $\text{top-}p=0.95$.
We sample $k=16$ completions per problem and compute pass@1, which provides more reliable performance estimates.

\textbf{Baseline Methods.} We compare \methodname against state-of-the-art KV-cache compression methods.
We adapt SnapKV~\cite{li2024snapkv} to operate in both the prefill and decoding phases.
We also evaluate {R-KV~\cite{rkv}}, a leading KV compression method for reasoning models, using its recommended configuration.
For all baselines, we apply a fixed compression schedule that recompresses the KV cache every 128 decoding steps.
The memory budget is defined over the entire KV cache, including segments from generated tokens that are subsequently compressed.
We also report FullKV, which keeps the KV cache uncompressed and serves as the gold-standard upper bound on decoding quality.
The budget for SnapKV and R-KV represents the number of KV cache entries, while for \methodname, we adopt 75\% entries in full precision and 25\% entries quantized.
When plotting, we normalize \methodname's budget to its memory-equivalent logical budget, allowing for a direct comparison against the full-precision baselines under the same actual memory consumption.

\begin{figure}[!t] %
    \centering
    
    \begin{subfigure}[b]{\linewidth}
        \centering
        \includegraphics[width=\linewidth]{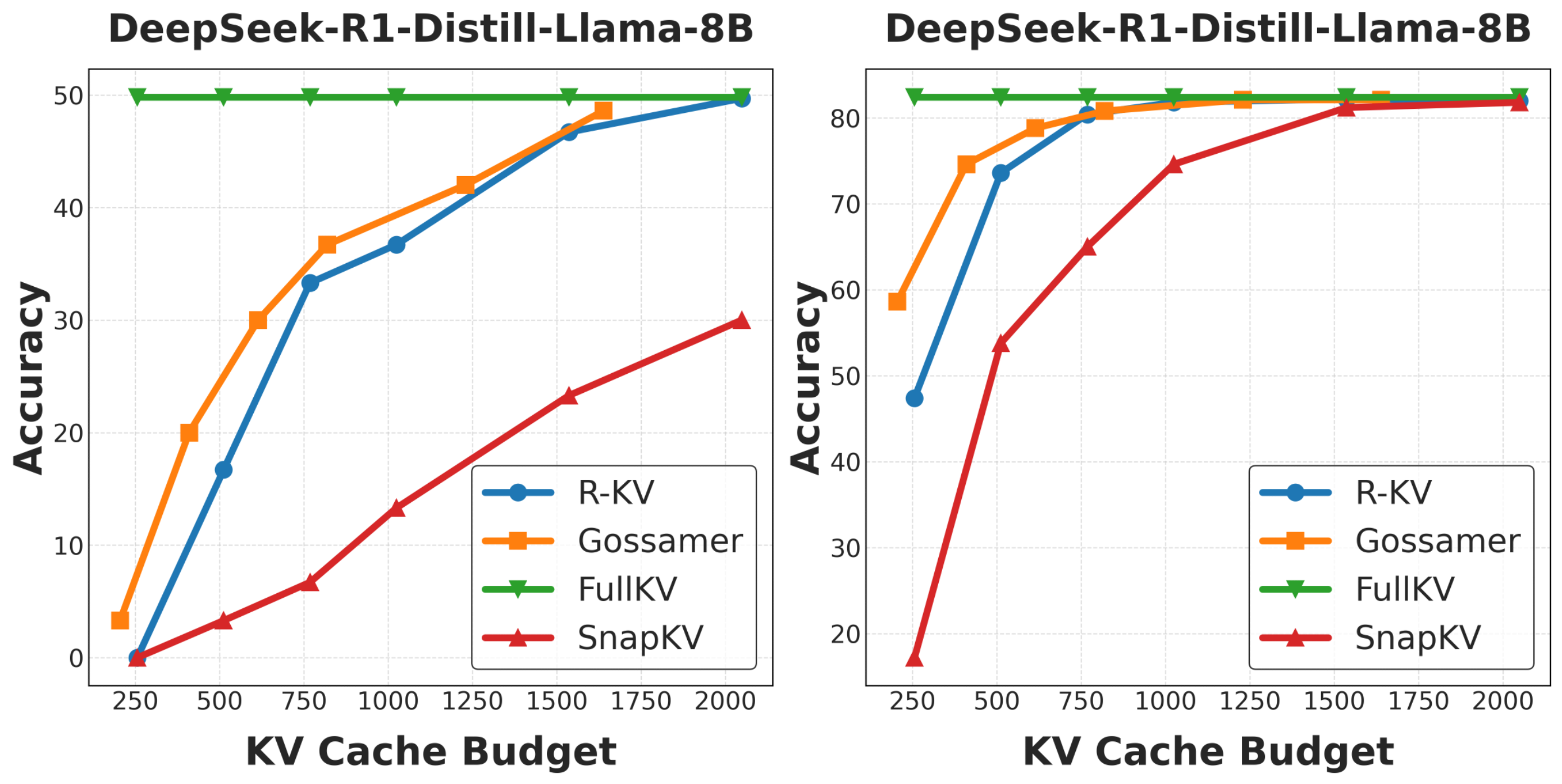}
        % \vspace{}
        \caption{{Accuracy on AIME (left) / MATH (right) for Llama-8b.}}
        % \caption{{Accuracy on AIME (left) / MATH (right) for Llama-8b vs. KV Cache Budgets}}
        \label{fig:6_1_1_accuracy_math}
    \end{subfigure}
    
    \vskip 0.5em  %
    
    \begin{subfigure}[b]{\linewidth}
        \centering
        \includegraphics[width=\linewidth]{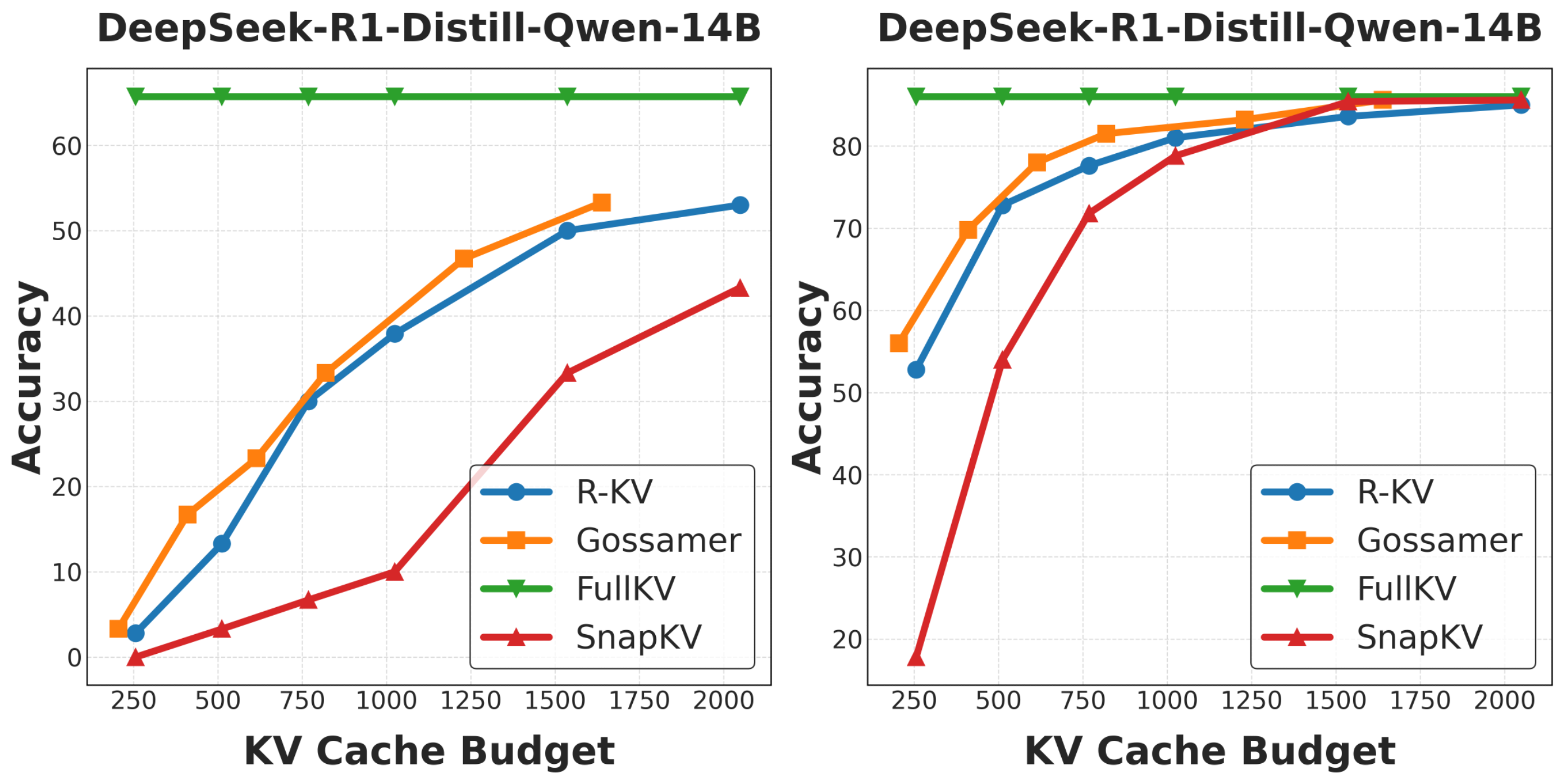}
        \caption{{Accuracy on AIME (left) / MATH (right) for Qwen-14b.}}
        % \caption{{Accuracy on AIME (left) / MATH (right) for Qwen-14b vs. KV Cache Budgets}}
        \label{fig:6_1_2_accuracy_aime24}
    \end{subfigure}

    \caption{
      % Accuracy vs. KV Budget Across Models and Datasets. AIME 2024 (left); Math-500 (right)
      Accuracy vs. KV Budget Across Models and Datasets.}
    \label{fig:6_1_accuracy}
    \vspace{-1.8em}
\end{figure}

\textbf{Accuracy Analysis.}
Across models and benchmarks, \methodname consistently stays closest to the FullKV upper bound while outperforming existing compression baselines (\cref{fig:6_1_accuracy}). This trend is especially clear on AIME 2024, where accuracy is highly sensitive to KV budget. 
For both DeepSeek-R1-Distill-Llama-8B and DeepSeek-R1-Distill-Qwen-14B, \methodname achieves higher accuracy than R-KV and SnapKV throughout the low- and mid-budget regimes, and closes the gap to FullKV at substantially smaller budgets. 
In other words, \methodname recovers the uncompressed accuracy frontier more quickly as KV capacity increases, indicating that it allocates limited cache budget more effectively than prior methods.

On MATH-500, the accuracy curves are smoother and the absolute gaps are smaller, but the relative ordering remains consistent. For Llama-8B, \methodname matches or slightly exceeds R-KV across all budgets while maintaining a clear advantage over SnapKV under tight memory (256--768). This suggests that \methodname makes more efficient use of constrained KV capacity, whereas less frequent compression updates can lead to wasted budget and slower accuracy recovery. As the budget increases, all compressed methods gradually approach the FullKV limit, but \methodname reaches this regime sooner, showing that its gains are not limited to aggressive compression alone. For Qwen-14B, the same pattern holds: \methodname remains closest to FullKV, consistently outperforms R-KV in the low- and mid-budget range, and avoids the severe degradation exhibited by SnapKV at small budgets.

Taken together, these results show that \methodname is not only stronger at a fixed KV budget, but also more budget-efficient: it delivers larger accuracy gains as KV capacity increases and reaches the FullKV regime at smaller budgets than prior compression methods.

\subsection{Performance Evaluation}

\textbf{Simulation Setup.} 
To evaluate architectural performance, we developed a cycle-accurate simulation framework based on SCALEsim-v2~\cite{samajdar2018scale}.
The simulator accepts layer-wise traces generated from specific models and datasets, enabling precise modeling of runtime behavior.
We model the \methodname architecture, which features a 128$\times$128 PE array.
The PEs utilize an output stationary dataflow and perform FP16 multiplication with FP32 accumulation.
The architecture includes a total of 64 MB of on-chip buffer, partitioned into 16 MB for inputs, 16 MB for weights, and 32 MB for outputs.
For off-chip memory, we model HBM2 providing a total bandwidth of 512 GB/s.
We implement the \methodname architecture in SystemVerilog and synthesize it using Synopsys Design Compiler under a commercial 28nm technology node with 500 MHz clock frequency for area and energy analysis.

\textbf{Models and Baselines.} 
We evaluate architectural performance using DeepSeek-R1-Distill-Llama-8B, DeepSeek-R1-Distill-Qwen-14B, and \Rebuttal{QwQ-32B.}
We compare our \methodname approach against \Rebuttal{four} baseline methods. First, we simulate FullKV, where the entire KV cache is stored uncompressed.
\Rebuttal{Second, we include Ecco~\cite{cheng2025ecco} and Oaken~\cite{oaken}, two state-of-the-art quantization-based frameworks that compress the KV cache to near-4-bit precision with on-the-fly encoding and decoding.}
Third, we select R-KV as it is a representative state-of-the-art compression method for reasoning models.
For our own \methodname architecture, we incorporate two configurations: one where 50\% of the KV cache is allocated to FP16 and another with 75\%.
All baselines (FullKV, Ecco, and R-KV) are simulated using the identical SCALEsim-v2 setup to ensure a fair and direct comparison.

\begin{figure}[t]
  \centering
  \includegraphics[width=\linewidth]{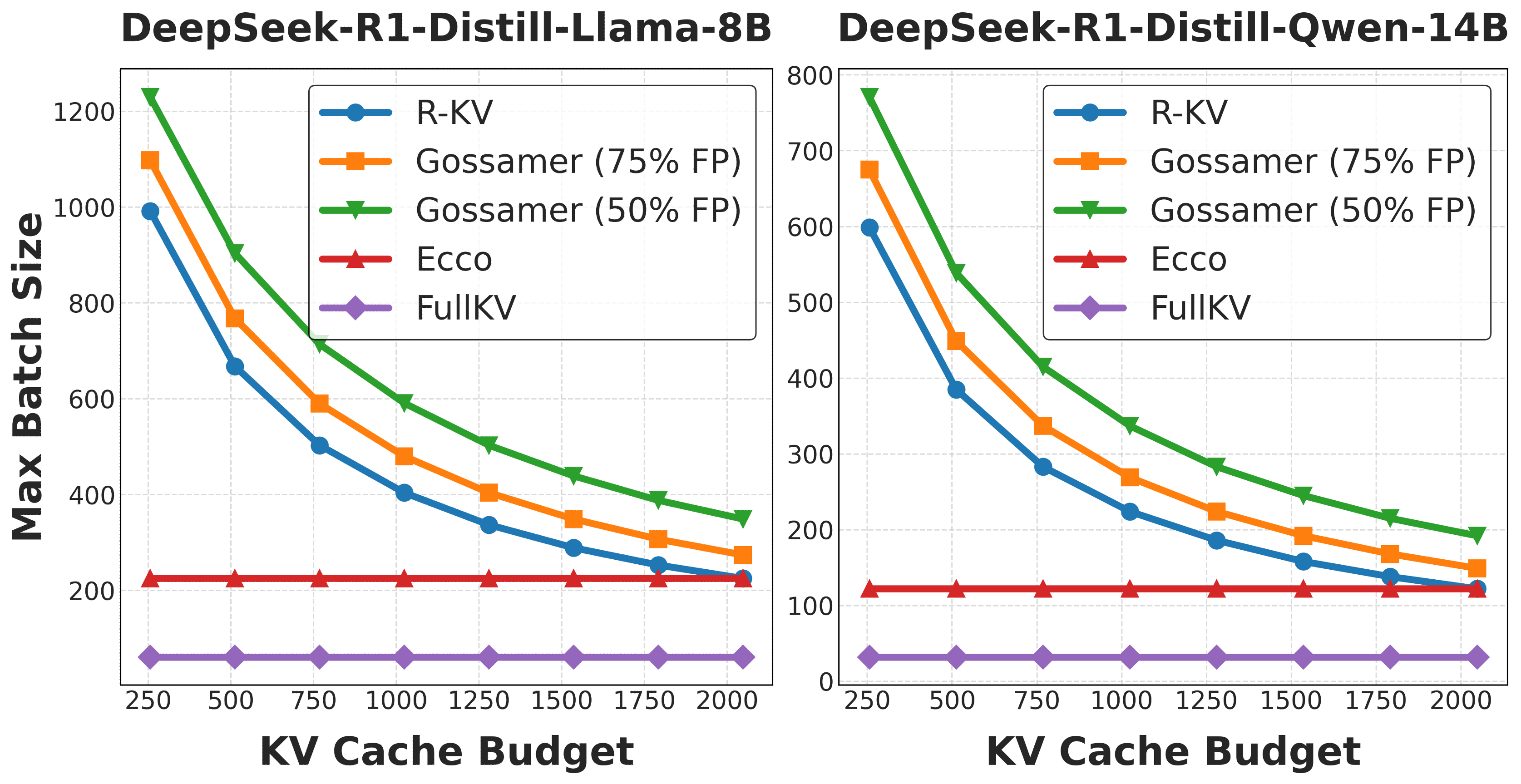}
  \caption{Max Batch Size vs. KV Cache Budget Across Models.}
  \label{fig:6_2_max_batch}
  \vspace{-1.8em}
\end{figure}

\begin{figure*}[!t] %
    \centering
    
    \begin{subfigure}[b]{\linewidth}
        \centering
        \includegraphics[width=\linewidth]{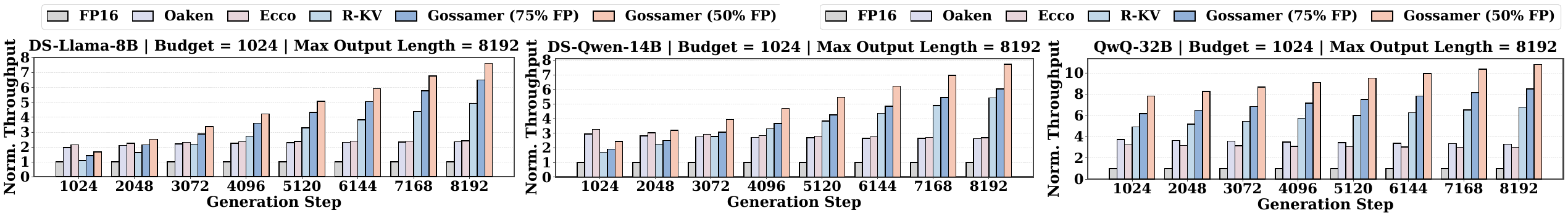}
        \caption{\Rebuttal{Throughput vs. generation step with KV cache budget\,=\,1024 for DS-Llama-8B (left), DS-Qwen-14B (middle), and QwQ-32B (right).}}
        \label{fig:6_3_1_single_step_llama}
    \end{subfigure}
    
    \vskip 0.5em  %
    
    \begin{subfigure}[b]{\linewidth}
        \centering
        \includegraphics[width=\linewidth]{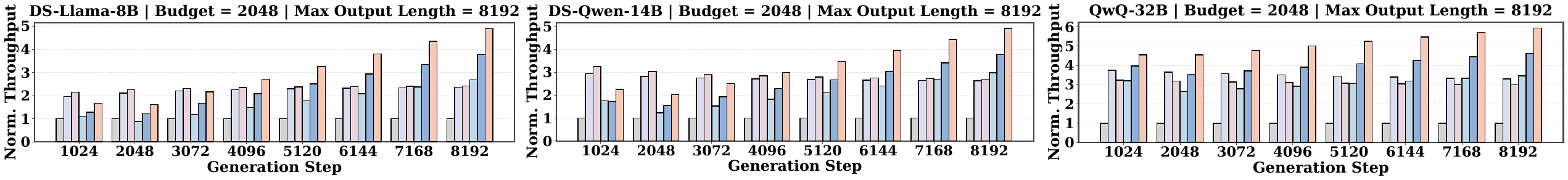}
        \caption{\Rebuttal{Throughput vs. generation step with KV cache budget\,=\,2048 for DS-Llama-8B (left), DS-Qwen-14B (middle), and QwQ-32B (right).}}
        \label{fig:6_3_2_single_step_qwen}
    \end{subfigure}
    
    \vskip 0.5em  %
    
    \begin{subfigure}[b]{\linewidth}
        \centering
        \includegraphics[width=\linewidth]{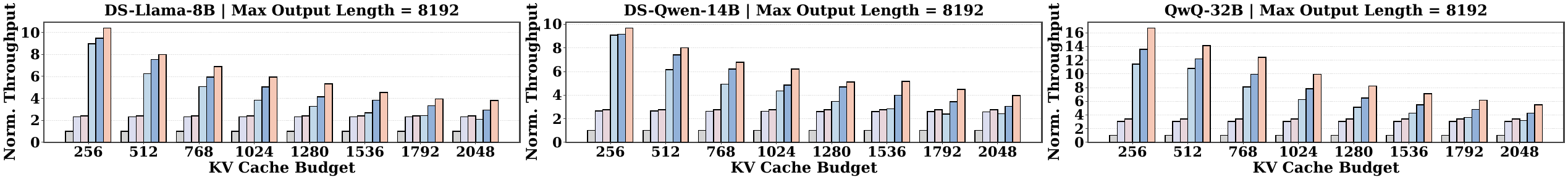}
        \caption{\Rebuttal{Normalized end-to-end throughput vs. KV cache budget for DS-Llama-8B (left), DS-Qwen-14B (middle), and QwQ-32B (right).}}
        \label{fig:6_3_3_e2e_model_across_budget}
    \end{subfigure}

    \caption{ \Rebuttal{
      Normalized throughput across varying generation steps and KV cache budgets.}
    }
    \label{fig:6_3_throughput}
    \vspace{-1.2em}
\end{figure*}

\textbf{Memory Saving and Max Batch Size.}
To measure the throughput benefit, we compute the maximum batch size that fits in device memory for each method and KV budget when generating up to 8,192 tokens per sequence. In this regime, KV cache is the dominant memory cost, so reducing its footprint directly increases the number of sequences that can be served concurrently. \cref{fig:6_2_max_batch} reports the resulting batch-size curves for both DeepSeek-R1-Distill-Llama-8B and DeepSeek-R1-Distill-Qwen-14B.

Across both models and all budgets, \methodname achieves the highest batch size among the compressed methods and substantially outperforms FullKV. The gain is largest in the low-budget regime, where memory is most constrained: the 50\%-FP variant supports more than an order-of-magnitude larger batch than FullKV at a budget of 256, and consistently exceeds R-KV by roughly 20--40\%. The 75\%-FP variant follows the same trend, offering a slightly smaller concurrency gain in exchange for better accuracy. As the budget increases, all methods gradually converge, but \methodname remains above R-KV and Ecco throughout, showing that its compression translates into systematically better memory efficiency rather than a benefit limited to only the most aggressive settings.

Overall, these results show that \methodname improves not only accuracy under tight KV budgets, but also effective hardware utilization by enabling substantially larger long-context batches on a fixed device.

\Rebuttal{
\textbf{Decoding Throughput.} 
\cref{fig:6_3_throughput}-a and \cref{fig:6_3_throughput}-b report normalized decoding throughput over generation steps at budgets of 1024 and 2048, across all three models.
All compressed methods eventually outperform FP16 FullKV as sequence length grows, since smaller KV footprints reduce memory traffic and attention cost. The key distinction is between the short-generation regime and the long-output regime typical of reasoning workloads. In early steps (KV length within budget), Ecco and Oaken lead due to immediate bandwidth savings from 4-bit quantization, while R-KV and \methodname show smaller initial gains as budget-based compression has not yet activated. Once KV length exceeds the budget and compression becomes active, \methodname becomes increasingly advantageous, ultimately delivering the highest throughput across all models and budgets.

The source of \methodname's speedup is threefold: increased batch size from reduced KV memory footprint, lower per-step attention cost from mixed-precision KV storage, and negligible compression overhead from fused similarity estimation. The relative contribution of each factor varies with model size. For smaller models (DS-Llama-8B, DS-Qwen-14B), GEMM operations saturate quickly at modest batch sizes, so attention becomes the dominant cost as sequence length grows — making KV compression increasingly impactful. For QwQ-32B, large weight matrices consume most of the device memory, leaving room for only small batches; in this regime, attention contributes a smaller fraction of total latency, so \methodname's throughput gain is primarily driven by the increased batch size that reduced KV footprint enables. This explains why QwQ-32B achieves the highest absolute normalized throughput among the three models, with \methodname (50\% FP) reaching over 10\x FP16 throughput at 8192 tokens. 

At budget $=2048$, the crossover between \methodname and quantization only baselines occurs later since the larger budget delays compression pressure, but the final advantage at 8,192 tokens remains consistent. The gap between \methodname and R-KV widens with sequence length across all models, confirming that the benefit of fused compression compounds as generation progresses.
}

\Rebuttal{
\textbf{End-to-end Analysis.}
\cref{fig:6_3_throughput}-c reports normalized end-to-end throughput over full 8192-token generation. \methodname consistently delivers the best performance across all three models and all budgets, achieving up to 16.7\x over FullKV, 5.5\x over Oaken, 4.9\x over Ecco, and 1.8\x over R-KV. Throughput decreases as budget grows for all methods, since a larger retained cache increases memory and attention cost; however, \methodname maintains a consistent lead over all baselines throughout the accuracy-relevant regime (1024--2048), where both the 75\%-FP and 50\%-FP variants sustain strong throughput while preserving reasoning quality. These results demonstrate that \methodname's advantages: increased batch capacity, mixed-precision attention, and fused compression, compound across model scales and translate into the highest end-to-end throughput under practical long-context serving conditions.
}

\begin{figure}[t]
  \centering
  \includegraphics[width=\linewidth]{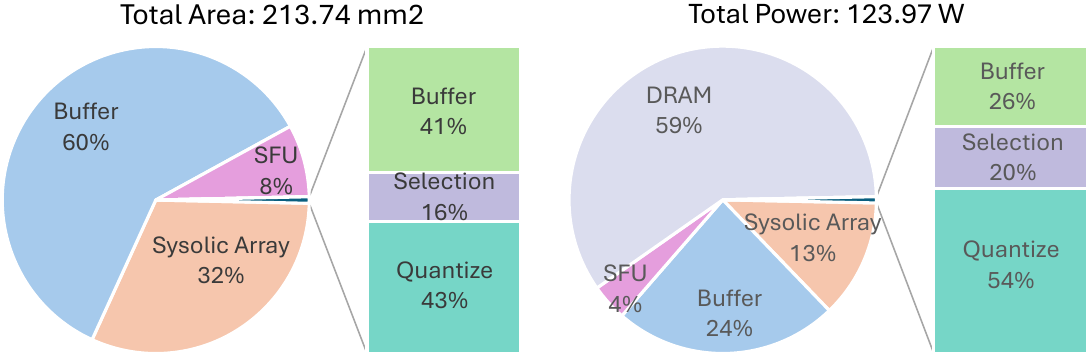}
  \caption{\Rebuttal{Area and Power Breakdown.}}
  \label{fig:power_area}
  \vspace{-1em}
\end{figure}

\Rebuttal{
\textbf{Power and Area Analysis.}
The area and power consumption of our proposed architecture \methodname are summarized in \cref{fig:power_area}. 
By employing a highly streamlined microarchitecture and leveraging idle compute cycles, \methodname requires minimal additional logic for token selection and quantization. Consequently, the hardware implementation is exceptionally lightweight. Synthesis results demonstrate that our custom module occupies just 1.46 mm² of on-chip area and consumes 811 mW of power. This translates to a negligible overhead of only 0.68\% in total area and 0.65\% in total power consumption relative to the baseline system.
}

\subsection{Ablation Study}

\begin{figure}[t]
  \centering
  \includegraphics[width=\linewidth]{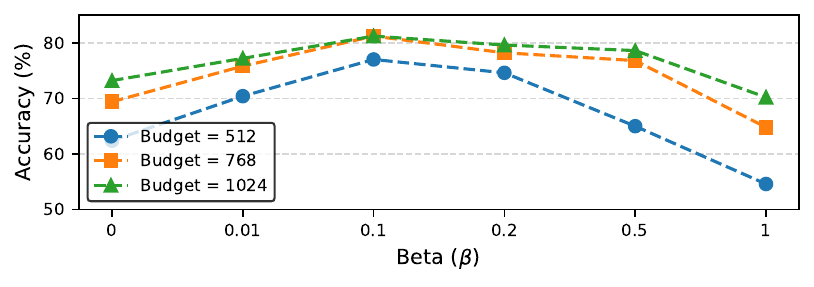}
  \vspace{-2.em}
  \caption{Redundancy-Importance trade-off parameter $\beta$.}
  \label{fig:beta_tradeoff}
  \vspace{-1em}
\end{figure}

\textbf{Redundancy-Importance Trade-off.} 
To determine token retention, we introduce a hybrid criterion that jointly evaluates token redundancy and importance, regulated by a weighting parameter, $\beta$. We evaluate the sensitivity of this trade-off by sweeping $\beta$ under different token budgets using the Llama-8B model on the MATH-500 dataset. As illustrated in \cref{fig:beta_tradeoff}, $\beta = 0$ represents a purely redundancy-based policy, while $\beta = 1$ relies strictly on token importance. As $\beta$ scales from 0 to 1, accuracy consistently exhibits an inverted-U trend across all budgets. Peak performance is universally achieved at $\beta = 0.1$, demonstrating that strategically combining both metrics yields significantly higher accuracy than relying on either isolated extreme.

\begin{figure}[t]
  \centering
  \includegraphics[width=\linewidth]{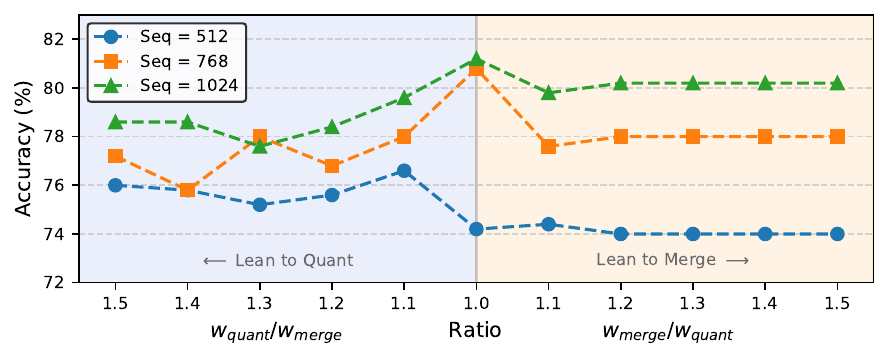}
  \vspace{-2.em}
  \caption{\Rebuttal{Accuracy vs. merge-quant decision threshold.}}
  \label{fig:w_tradeoff}
  \vspace{-1.5em}
\end{figure}

\Rebuttal{
\textbf{Merge-Quantization Sensitivity.} 
\cref{fig:w_tradeoff} sweeps the ratio between $w_\text{merge}$ and $w_\text{quant}$, where values left of center bias decisions toward quantization and values right toward merging.
For larger budgets, 768 and 1024, accuracy peaks precisely at a ratio of 1.0. This indicates that when the budget is sufficiently large, the system is best served by an unbiased decision that simply selects the method yielding the minimum information loss. Conversely, under a highly constrained budget 512, the peak shifts leftward to a $w_\text{quant}/w_\text{merge} = 1.1$. At this aggressive compression level, the available FP capacity is scarce, meaning almost all retained tokens are highly critical. Forcing a merge operation on these critical tokens risks severe context distortion. Thus, an explicit bias toward quantization is required to maintain accuracy under strict budget constraints.
}

\begin{figure}[t]
  \centering
  \includegraphics[width=\linewidth]{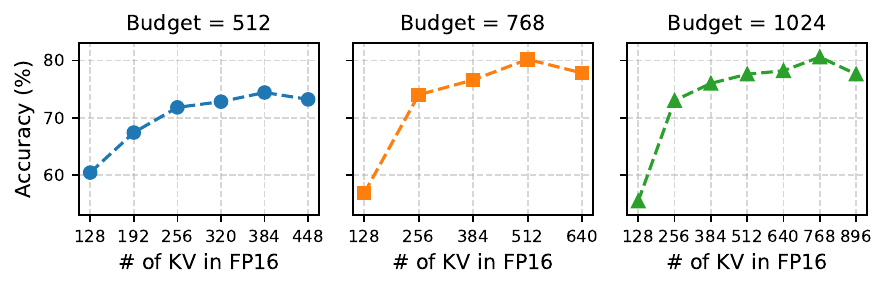}
  \vspace{-2.em}
  \caption{Accuracy vs. the number of FP16 KV cache entries.}
  \label{fig:fp_tradeoff}
  \vspace{-1.2em}
\end{figure}

\textbf{Budget Partitioning Analysis.}
\cref{fig:fp_tradeoff} sweeps the FP16 allocation under different total budgets. Accuracy generally improves with more FP entries, with peak performance consistently achieved at 50\%--75\% FP allocation.

Two non-obvious effects emerge. First, at a severely constrained FP allocation of 128 entries, the smallest budget (512) outperforms larger ones (768, 1024): with more total tokens but the same FP capacity, larger budgets force more tokens into aggressive quantization, distorting the contextual representation. This suggests quantization should be applied selectively rather than indiscriminately. Second, allocating nearly the entire budget to FP does not always help: under budget = 1024, accuracy slightly drops when increasing FP from 768 to 896 entries, indicating that moderate quantization can act as a regularizer by filtering contextual noise.

\begin{figure}[t]
  \centering
  \includegraphics[width=\linewidth]{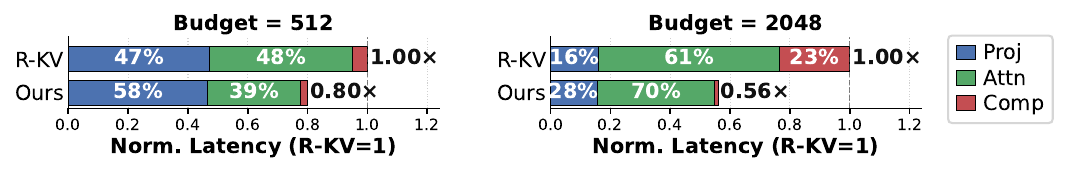}
  \vspace{-1.5em}
  \caption{\Rebuttal{Normalized execution latency breakdown.}}
  \label{fig:latency_breakdown}
  \vspace{-1em}
\end{figure}

\Rebuttal{
\textbf{Execution Time Breakdown.}  
\cref{fig:latency_breakdown} breaks down Llama-8B end-to-end execution latency into three components, projection (Proj), attention (Attn), and compression (Comp), normalized to R-KV at each budget.
Projection latency remains roughly constant in absolute time across both methods. Attention latency is shorter in absolute terms under Gossamer because mixed-precision KV storage reduces the amount of data streamed from memory per decoding step, directly lowering the memory-bound attention cost.

The most significant difference lies in compression overhead. In R-KV, compression requires a dedicated global similarity pass whose cost scales with sequence length; at budget = 2048, this accounts for 23\% of total runtime. By fusing cosine similarity estimation into the attention GEMM, Gossamer reduces compression overhead to under 2\% across all budgets, effectively hiding it within the attention kernel. This is the core benefit of Free-Lunch Utilization: compression that would otherwise dominate at large budgets is absorbed into otherwise idle compute cycles. The combined effect yields a 1.25\x speedup over R-KV at budget = 512 and a 1.79\x speedup at budget = 2048, with the advantage growing as budget increases, since R-KV's compression cost scales while Gossamer's remains negligible.
}

\Rebuttal{
\textbf{Long-Context Capability.} 
\Cref{tab:evaluation:longbench} evaluates Gossamer on LongBench v1 and v2 under a strict KV cache budget of 1024 and 2048 tokens, respectively. On LongBench v1 (Llama-3.1-8B-Instruct), Gossamer matches or exceeds FullKV on MF-en and Qasp while remaining competitive on Musiq, consistently surpassing KIVI, Quest and Cake by a significant margin. On LongBench v2 (DS-Llama-8B), Gossamer achieves 30.4\% overall accuracy across short and medium length categories, within 0.4\% of FullKV and outperforming all compressed baselines. These results demonstrate that Gossamer's hybrid compression strategy generalizes effectively beyond reasoning workloads to diverse long-context tasks.
}

\Rebuttal{
\textbf{Cross-Domain Codebook Adaptivity.}
Notably, all LongBench evaluations reuse the quantization codebook profiled exclusively on GSM8K — a mathematical reasoning dataset with a very different token distribution. Despite this domain shift, Gossamer maintains high accuracy across varied long-context tasks, demonstrating zero-shot codebook transferability. This eliminates the need for task-specific KV cache re-profiling and significantly simplifies deployment.
}

\begin{table}[!t]
\centering
\caption{{LB: DS-Llama-8B and Llama-3.1-8B-Instruct.}}
\begin{tabular}{l|c|cc|ccc}
\toprule
  & \multicolumn{3}{c|}{Longbench v2} &
\multicolumn{3}{c}{Longbench v1}  \\
  & \multicolumn{3}{c|}{Deepseek-Llama-8B} &
\multicolumn{3}{c}{Llama-3.1-8B-Instruct}  \\
\midrule
Method  
&  Acc.    &   Short   &   Medium &  Qasp    &   MF-en   &   Musiq \\
\midrule
FullKV  
&   30.8\%    &   36.1\%  &   26.3\%     &   45.31    &   54.97  &   31.68     \\
KIVI \cite{liu2024kivi}
& -	&   -	   & -       & 36.50	&   42.51	   & 25.52     \\
Quest \cite{tang2024quest}
& 27.3\%	&   30.2\%	   & 24.9\%       & 40.22	&   48.62	   & 26.78     \\
Cake \cite{qin2025cake}
& 29.2\%	&    33.6\%	   & 25.6\%      & 44.95	&    52.48	& 30.82        \\ 
\textbf{Ours} 
& \textbf{30.4\%}	   &    \textbf{35.8\%}	   &   \textbf{25.9\%} & \textbf{45.19}	   &    \textbf{55.24}	   &   \textbf{30.21} \\
\bottomrule
\end{tabular}
\label{tab:evaluation:longbench}
\vspace{-1em}
\end{table}

\begin{figure}[t]
  \centering
  \includegraphics[width=\linewidth]{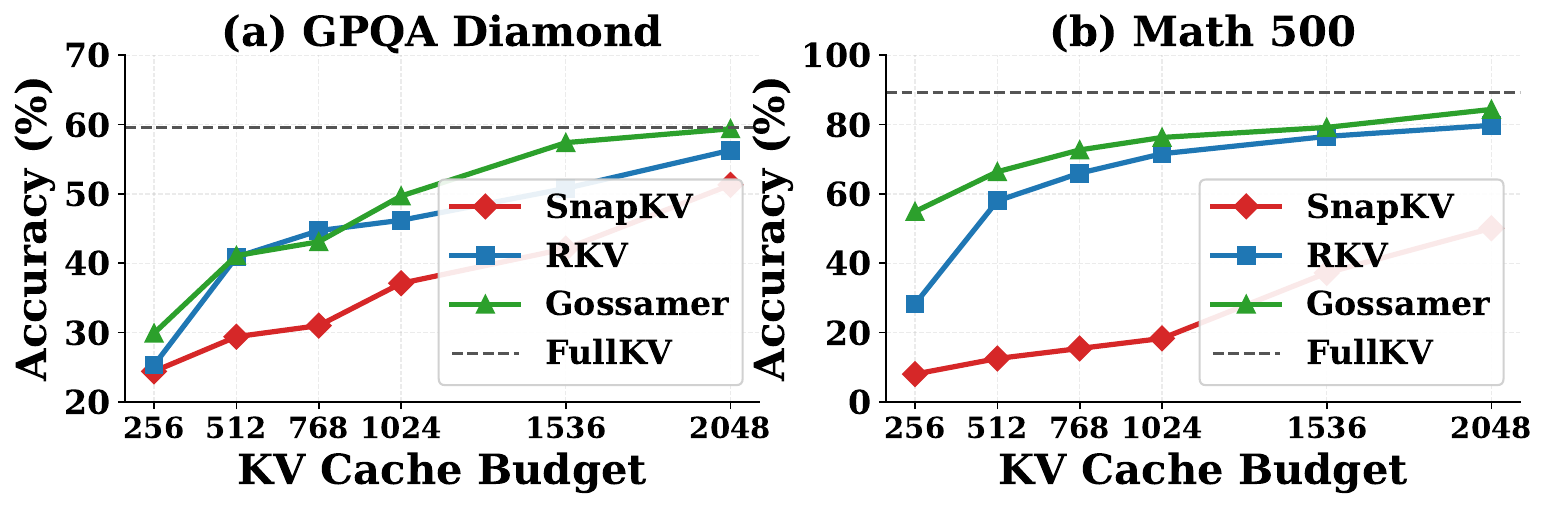}
  \caption{\Rebuttal{{GPQA/MATH acc. vs. KV cache budget (QwQ-32B).}}}
  \label{fig:qwq_accu}
  \vspace{-1.2em}
\end{figure}

\Rebuttal{
\textbf{Accuracy Scalability.}
\cref{fig:qwq_accu} evaluates Gossamer on QwQ-32B, a larger and stronger reasoning model, across two benchmarks: GPQA Diamond and MATH-500. Gossamer consistently outperforms both baselines across nearly all budgets on both tasks. The advantage is most pronounced at tight budgets: on MATH-500 at budget = 256, Gossamer achieves 55.0\% accuracy, nearly doubling R-KV and far exceeding SnapKV; a similar gap appears on GPQA Diamond, where Gossamer leads both baselines at low budgets while SnapKV trails substantially throughout. As the budget increases, all methods converge toward FullKV, with Gossamer remaining at or above R-KV on both benchmarks and consistently ahead of SnapKV. These results confirm that Gossamer's accuracy gains scale to larger models and generalize across reasoning task types, rather than being limited to the smaller distilled checkpoints and single benchmark used in the main evaluation.
}

\Rebuttal{
\textbf{Action Distribution Analysis.} 
To verify that Gossamer's gains stem from genuinely leveraging all three actions rather than degenerating into one dominant policy, we record the fraction of candidate tokens assigned to each action by the DAU framework. Under budget = 1024 on MATH-500 with DS-Llama-8B, the distribution is: Evict 87.6\%, Quantize 10.1\%, and Merge 2.3\%.

This distribution confirms that no single action subsumes the others — removing quantization or merging would force their respective tokens into eviction, discarding information the DAU framework otherwise preserves at low cost. This supports our central claim that a hybrid, utility-driven action space, rather than any single compression technique, is necessary to make effective use of a constrained KV budget.
}

\section{Conclusion}
\label{sec:conclusion}
This paper presented Gossamer, a utility-driven KV cache compression architecture that addresses memory and bandwidth bottlenecks in long-form reasoning LLMs.
By unifying quantization, merging, and eviction under a constant-budget framework and fusing utility estimation into attention computation, Gossamer achieves predictable memory usage with negligible overhead.
Evaluations demonstrate that Gossamer sustains reasoning quality comparable to uncompressed inference while delivering up to 16.7\x throughput improvement over FP16 baselines and significant gains over state-of-the-art compression methods.

\section*{Acknowledgment}
The work was funded in part by National Science Foundation NSF-2112562 and ARO W911NF-23-2-0224. The authors sincerely thank the anonymous reviewers for their constructive feedback and valuable suggestions that greatly improved the quality of this work.

% \balance
\bibliographystyle{IEEEtranS}
\bibliography{sample-base}

\end{document}